\documentclass{emulateapj}

\usepackage{graphics,graphicx}
\usepackage{subfigure}
\usepackage{color}
\usepackage{placeins}
\usepackage{tabu}

\begin{document}

\title{The 31 Deg$^2$ Release of the Stripe 82 X-ray Survey: The Point Source Catalog}

\author{Stephanie M. LaMassa$^{1,2,3}$,
C. Megan Urry$^{1,2}$,
Nico Cappelluti$^4$,
Hans B\"{o}hringer$^{5}$,
Andrea Comastri$^4$,
Eilat Glikman$^6$,
Gordon Richards$^7$,
Tonima Ananna$^{1,2}$,
Marcella Brusa$^{4,8}$,
Carie Cardamone$^{9}$,
Gayoung Chon$^5$,
Francesca Civano$^{1,10}$,
Duncan Farrah$^{11}$,
Marat Gilfanov$^{12,13}$,
Paul Green$^{10}$,
S. Komossa$^{14}$,
Paulina Lira$^{15}$,
Martin Makler$^{16}$,
Stefano Marchesi$^{1,4,10}$,
Robert Pecoraro$^{1,2}$,
Piero Ranalli$^{17,4}$,
Mara Salvato$^5$,
Kevin Schawinski$^{18}$,
Daniel Stern$^{19}$,
Ezequiel Treister$^{20}$,
Marco Viero$^{21,22}$
}

\affil{$^1$Yale Center for Astronomy \& Astrophysics, Physics Department, P.O. Box 208120, New Haven, CT 06520, USA,
$^2$Department of Physics, Yale University, P.O. Box 208121, New Haven, CT 06520, USA, 
$^3$Now at NASA Goddard Space Flight Center, Greenbelt, MD 20771, USA,
$^4$INAF-Osservatorio Astronomico di Bologna, via Ranzani 1, I-40127 Bologna, Italy,
$^5$Max-Planck-Institut f\"ur extraterrestrische Physik, 85748 Garching, Germany,
$^6$Department of Physics, Middlebury College, Middlebury, VT 05753, USA
$^7$Department of Physics, Drexel University, 3141 Chestnut Street, Philadelphia, PA 19104, USA,
$^8$DIFA—Dipartimento di Fisica e Astronomia, Universita' di Bologna, viale Berti Pichat 6/2, I-40127 Bologna, Italy, 
$^9$Department of Math \& Science, Wheelock College, 200 Riverway, Boston, Massachusetts 02215,
$^{10}$Smithsonian Astrophysical Observatory, 60 Garden Street, Cambridge, MA 02138, USA,
$^{11}$Department of Physics MC 0435, Virginia Polytechnic Institute and State University, 850 West Campus Drive, Blacksburg, VA 24061, USA,
$^{12}$Max-Planck Institut f\"ur Astrophysik, Karl-Schwarzschild-Str. 1, Postfach 1317, D-85741 Garching, Germany
$^{13}$Space Research Institute of Russian Academy of Sciences, Profsoyuznaya 84/32,117997 Moscow, Russia,
$^{14}$Max-Planck-Institut f\"ur Radioastronomie, Auf dem H\"ugel 69, D-53121 Bonn, Germany,
$^{15}$Departamento de Astronomia, Universidad de Chile, Camino del Observatorio 1515, Santiago, Chile,
$^{16}$Centro Brasileiro de Pesquisas F\'isicas, Rua Dr Xavier Sigaud 150, Rio de Janeiro, RJ 22290-180, Brazil,
$^{17}$Institute for Astronomy, Astrophysics, Space Applications and Remote Sensing (IAASARS), National Observatory of Athens, 15236 Penteli, Greece,
$^{18}$Institute for Astronomy, Department of Physics, ETH Zurich, Wolfgang-Pauli Strasse 27, CH-8093 Zurich, Switzerland,
$^{19}$Jet Propulsion Laboratory, California Institute of Technology, Pasadena, CA 91109, USA,
$^{20}$Department of Astronomy, University of Concepcion, Concepcion, Chile,
$^{21}$Kavli Institute for Particle Astrophysics and Cosmology, Stanford University, 382 Via Pueblo Mall, Stanford, CA 94305, USA; California Institute of Technology, 1200 E. California Blvd., Pasadena, CA 91125, USA,
$^{22}$California Institute of Technology, 1200 E. California Blvd., Pasadena, CA 91125, USA
}

\begin{abstract}
We release the next installment of the Stripe 82 X-ray survey point-source catalog, which currently covers 31.3 deg$^2$ of the Sloan Digital Sky Survey (SDSS) Stripe 82 Legacy field. In total, 6181 unique X-ray sources are significantly detected with {\it XMM-Newton} ($>5\sigma$) and {\it Chandra} ($>4.5\sigma$). This catalog release includes data from {\it XMM-Newton} cycle AO 13, which approximately doubled the Stripe 82X survey area. The flux limits of the Stripe 82X survey are $8.7\times10^{-16}$ erg s$^{-1}$ cm$^{-2}$, $4.7\times10^{-15}$ erg s$^{-1}$ cm$^{-2}$, and $2.1\times10^{-15}$ erg s$^{-1}$ cm$^{-2}$ in the soft (0.5-2 keV), hard (2-10 keV), and full bands (0.5-10 keV), respectively, with approximate half-area survey flux limits of $5.4\times10^{-15}$ erg s$^{-1}$ cm$^{-2}$, $2.9\times10^{-14}$ erg s$^{-1}$ cm$^{-2}$, and $1.7\times10^{-14}$ erg s$^{-1}$ cm$^{-2}$. We matched the X-ray source lists to available multi-wavelength catalogs, including updated matches to the previous release of the Stripe 82X survey; 88\% of the sample is matched to a multi-wavelength counterpart. Due to the wide area of Stripe 82X and rich ancillary multi-wavelength data, including coadded SDSS photometry, mid-infrared {\it WISE} coverage, near-infrared coverage from UKIDSS and VHS, ultraviolet coverage from {\it GALEX}, radio coverage from FIRST, and far-infrared coverage from {\it Herschel}, as well as existing $\sim$30\% optical spectroscopic completeness, we are beginning to uncover rare objects, such as obscured high-luminosity AGN at high-redshift. The Stripe 82X point source catalog is a valuable dataset for constraining how this population grows and evolves, as well as for studying how they interact with the galaxies in which they live.
\end{abstract}


\section{Introduction}
Active galactic nuclei (AGN) signal the growth of supermassive black holes at galactic centers. Studying AGN over a range of redshift allows us to discover how supermassive black holes evolve over cosmic time to the present day. As AGN emit energy over a range of wavelengths, they can be identified by various signatures, including optical and ultraviolet light from the accretion disk feeding the black hole \citep{koratkar}, optical emission from gas ionized by accretion disk photons \citep{bpt,vo,vandenberk}, X-ray emission from the AGN corona \citep{haardt,brandt2015}, mid-infrared emission from AGN heated circumnuclear dust \citep{spinoglio,lacy,stern2005,donley,stern2012,assef} and fine-structure emission lines \citep{farrah2007,melendez,weaver}, and radio emission from jets launched by the accretion disk \citep{kellermann,hooper}. These different selection criteria favor different parts of the AGN population, and by combining these methods, a comprehensive view of black hole growth is revealed.

Multi-wavelength surveys are then the key for unlocking the secrets of AGN evolution and how they relate to the galaxies they inhabit. Complementary survey strategies select different populations in the redshift-luminosity plane. Deep, pencil-beam surveys uncover the faintest objects in the Universe while wide-area surveys are required to discover a representative sampling of rare objects that have a low space density. Such rare sources include high-luminosity AGN at high-redshift (e.g., $L_{x} > 10^{45}$ erg s$^{-1}$ at $z>2$), which according to current theories, are the phase when most of the mass locked up in current black holes was accreted \citep[e.g.,][]{hopkins2009,treister2012}.

Wide-area surveys have existed for years at optical, infrared, and radio wavelengths, but have only recently been underway in X-rays at energies above 2 keV and at depths capable of pushing to cosmological distances. While the deep, small area {\it Chandra} Deep Field South Survey \citep[0.13 deg$^2$;][]{giacconi,xue} has uncovered the faintest AGN and has entered the flux regime where the number density of non-active galaxies surpasses that of active systems \citep{lehmer2012}, and medium-area surveys like {\it XMM-Newton} and {\it Chandra}-COSMOS \citep[2.2 deg$^2$;][]{xmm_hasinger,cappelluti,elvis2009,brusa2010,civano,civanoxvp,marchesi} have identified nearly 2,000 moderate-luminosity AGN ($10^{43}$ erg s$^{-1}< L_{x} < 10^{44}$ erg s$^{-1}$), the $L_{x} > 10^{45}$ erg s$^{-1}$ population has been a missing tier in our hard X-ray census of supermassive black hole growth. This population began to be revealed with larger area ($\sim$10 deg$^2$) surveys, such as XBo\"{o}tes \citep[9 deg$^{2}$;][]{murray,kenter} and the {\it Chandra} Multi-wavelength Project \citep[ChaMP, 10 deg$^2$;][]{kim2007}, as well as the more recent {\it XMM-Newton} survey in the {\it Herschel} ATLAS field \citep[7.1 deg$^2$;][]{hatlas}. The advent of the widest-area surveys ($>$15 deg$^2$), including the ``Stripe 82X'' survey \citep{me1,me2}, which as we discuss below, now reaches $\sim$31.3 deg$^2$, as well as the 50 deg$^2$ {\it XMM}-XXL (PI: Pierre)  and the $\sim$877 deg$^2$ {\it XMM}-Serendipitous \citep{xmm_seren5} surveys, provides a chance to study the evolution of the most luminous AGN in unprecedented detail. However, though the {\it XMM}-Serendipitous survey covers an order of magnitude more area than the dedicated large-area {\it XMM-Newton} surveys, an important component is missing: supporting multi-wavelength data which allows the X-ray photons to be identified with discrete sources and the properties of these objects to be characterized. A field which contains such supporting information, such as the Sloan Digital Sky Survey \citep[SDSS;][]{york} Stripe 82 region, is therefore an ideal location to execute an X-ray survey with maximal efficiency for returning comprehensive results.

Stripe 82 is a 300 deg$^2$ equatorial region imaged between 80 and 120 times as part of a supernova survey with SDSS \citep{frieman}. The coadded photometry reaches 1.2-2.2 magnitudes deeper than any single SDSS scan \citep[$r\sim24.6$ versus $r\sim22.2$;][]{annis,jiang}, and the full field has existing optical spectroscopy from SDSS and SDSS BOSS \citep[Data Releases 9 and 10;][]{ahn,dr10}, 2 SLAQ \citep{croom},  and WiggleZ \citep{drinkwater}, with partial coverage from DEEP2 \citep{newman}, PRIMUS \citep{coil}, 6dF \citep{jones2004,jones2009}, the VIMOS VLT Deep Survey \citep[VVDS][]{garilli}, a deep spectroscopic survey of faint quasars from \citet{jiang2006}, and a pre-BOSS pilot survey using Hectospec on MMT \citep{ross_mmt}. Existing multi-wavelength data in Stripe 82 include near-infrared observations from UKIDSS \citep{hewett,lawrence,casali} and the VISTA Hemisphere Survey \citep[VHS;][]{mcmahon}; far-infrared coverage from {\it Herschel} over 79 deg$^2$ \citep{viero}; ultraviolet coverage with {\it GALEX} \citep{morrissey}; radio observations at 1.4 GHz with FIRST \citep{becker1995,white,becker2012,helfand}, with deeper VLA coverage over 80 deg$^2$ \citep{hodge}; and millimeter observations with the Atacama Cosmology Telescope \citep[ACT;][]{fowler,swetz}. Additionally, there is {\it Spitzer} coverage in the field from the {\it Spitzer}-HETDEX Exploratory Large Area survey over 28 deg$^2$ (SHELA; PI: C. Papovich) and the {\it Spitzer} IRAC Equatorial Survey over 110 deg$^2$ (SpIES; PI: G. Richards; Timlin et al. {\it in prep.}),  deeper near-infrared $J$ and $K$ band coverage, to limits of 22 mag (AB), from the VISTA-CFHT Stripe 82 Survey over 140 deg$^2$ (VICS82,  PIs: Geach, Lin, Makler; J. Geach et al., {\it in prep.}), and mid-infrared coverage from the all-sky {\it WISE} mission \citep{wright}.

\begin{figure}
\centering
\includegraphics[angle=90,scale=0.4]{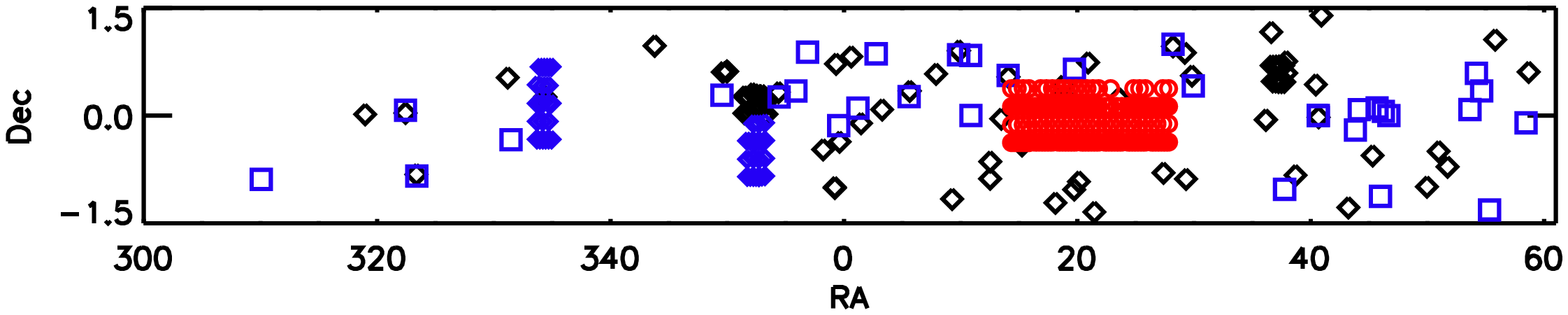}
\includegraphics[angle=90,scale=0.4]{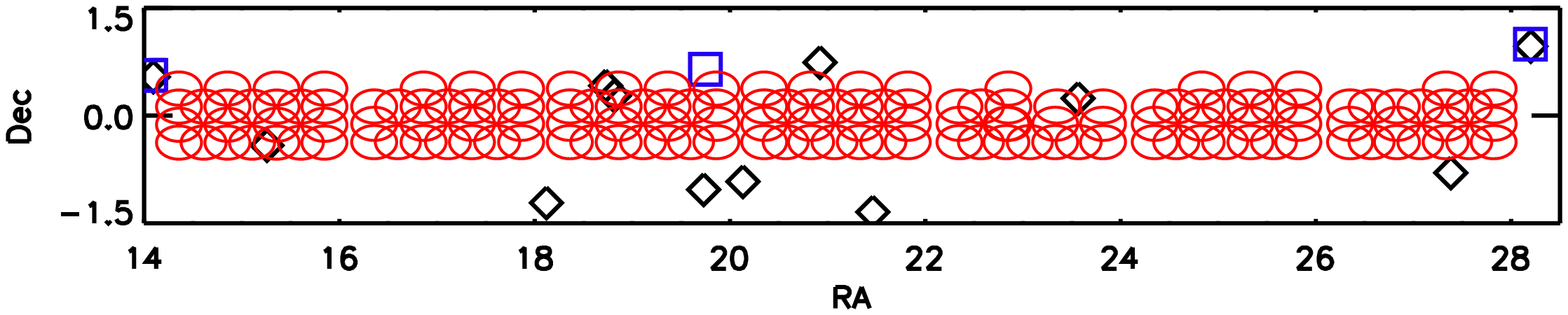}
\caption{\label{pointings} Distribution of archival {\it Chandra} observations (black diamonds), archival {\it XMM-Newton} observations (blue squares), {\it XMM-Newton} AO10 observations (blue diamonds), and {\it XMM-Newton} AO13 observations (red circles) for the full Stripe 82 region ({\it top}) and the {\it XMM-Newton} AO13 area ({\it bottom}). The symbol size is to scale with the field-of-view for the AO13 pointings in the bottom panel only.}
\end{figure}

To take advantage of this rich multi-wavelength coverage, we designed the wide-area Stripe 82X survey \citep{me1,me2} to uncover a representative population of rare, high-luminosity AGN at high redshift. Here we release the next installment of the Stripe 82X point-source catalog, which includes data awarded to our team in response to  {\it XMM-Newton} Announcement Opportunity 13 (``AO13''), representing $\sim$980 ks of observing time (PI: C. M. Urry; Proposal ID 074283). We also publish updated catalogs from our previous Stripe 82X data releases from archival {\it Chandra} and {\it XMM-Newton} observations in Stripe 82 \citep{me1,me2} and a pilot {\it XMM-Newton} program granted to our team in AO10 \citep[PI: C. M. Urry;][]{me2}. The positions of the X-ray pointings used in Stripe 82X are shown in Figure \ref{pointings}.

 In Section \ref{ao13_obs}, we discuss the data analysis for {\it XMM-Newton} AO13, which we then combine with the previously released {\it Chandra} and {\it XMM-Newton} data in Section \ref{s82x_prop} to characterize the Stripe 82 X-ray survey to date, currently spanning $\sim$31.3 deg$^2$ of non-overlapping area. In Section \ref{multi_wav}, we match the X-ray source lists to publicly available catalogs from SDSS, {\it WISE}, UKIDSS, VHS, {\it GALEX}, FIRST, and {\it Herschel}. Throughout, we adopt a cosmology of $H_{0}$ = 70 km s$^{-1}$ Mpc$^{-1}$, $\Omega_M = 0.27$ and $\Lambda=0.73$.

\section{{\it XMM-Newton} AO13 Observations and Data Analysis}\label{ao13_obs}
Our {\it XMM-Newton} AO13 program was executed between 2014 July and 2015 January in a series of seven observations, as summarized in Table \ref{xmm_summary}. Each observation consists of 22 individual pointings, or pseudo-exposures, which were carried out in ``mosaic mode.'' This observing mode efficiently surveys a large area with individual pointings that have relatively short exposure times. To reduce overhead, the EPIC offset tables are only uploaded (for the MOS detectors) and calculated (for the PN detector) for the first pointing in the series. In our AO13 program, each pseudo-exposure is separated by a half field-of-view ($\sim15^{\prime}$) to enable a greater depth to be achieved in the overlapping regions. The median exposure time for individual pointings before filtering is $\sim$5.2 ks for MOS1 and MOS2 and $\sim$4.7 ks for PN, while the coadded depth in the overlapping observations reaches $\sim$6-8 ks after filtering and correcting for vignetting (i.e., the energy-dependent decrease in effective area with off-axis distance).

The observational data files (ODF) were generated using the Science Analysis System (SAS) tasks (HEASOFT v. 6.16) {\it emproc} and {\it epproc} to create the MOS1, MOS2, PN, and PN out-of-time (OoT) events files. The OoT events occur from photons that are detected during CCD readout and recorded at random positions along the readout column. This effect is most significant for the PN detector and affects $\sim$6.3\% of the observing time. The PN images can be statistically corrected for this effect using the PN OoT files. 

The mosaicked observations were separated into individual pseudo-exposures using the Science Analysis System (SAS) package {\it emosaic\_prep}. Each pseudo-exposure was then filtered as described below.

\begin{deluxetable*}{lllllr}
\tablewidth{0pt}
\tablecaption{\label{xmm_summary} {\it XMM-Newton} AO13 Observation Summary}
\tablehead{\colhead{ObsID} & \colhead{Observation} & \colhead{Center RA} & \colhead{Center Dec} & 
  \colhead{Discarded} & \colhead{Area} \\ 
& \colhead{Date} & & & \colhead{Pseudo-exposures} & \colhead{(deg$^2$)}}

\startdata

0742830101 & 2014 Jul & 00:57:23.99 & -00:22:30.0 & \nodata & 2.33 \\

0747390101 & 2014 Jul & 01:05:23.99 & -00:22:30.0 & 22 (PN,M1,M2) & 2.22 \\

0747400101 & 2014 Jul & 01:13:24.00 & -00:22:30.0 & \nodata & 2.33 \\

0747410101 & 2015 Jan & 01:21:24.00 & -00:22:30.0 & 6 (PN), 8 (PN), 13 (PN) & 2.32 \\

0747420101 & 2015 Jan & 01:29:23.99 & -00:22:30.0 & 16 (PN,M1,M2), 18 (PN,M1,M2) & 1.95 \\
                     &                & & & 20 (PN,M1,M2), 21 (PN,M1,M2) \\

0747430101 & 2014 Jul & 01:37:23.99 & -00:22:30.0 & 22 (PN,M1,M2) & 2.22 \\

0747440101 & 2014 Aug & 01:45:23.99 & -00:22:30.0 & 22 (PN,M1,M2) & 2.22

\enddata

\end{deluxetable*}

\subsection{Flare Filtering}
Episodes of high levels of background radiation cause flaring in the {\it XMM-Newton} events files, hampering signal detection amidst the noise. To create good time intervals (GTIs), i.e., selecting events from observation periods where flaring is minimal, we started with a statistical approach. We created histograms of the count rate at high energies, 10-12 keV for the MOS detectors and 10-14 keV for the PN detector, in time bins of 100s, extracted from single events (PATTERN == 0). We created GTIs by excluding periods where the count rate was $\geq3\sigma$ above the mean and applied this filtering to the events file. From this events file, we searched for periods of low-energy (0.3-10 keV) flares, created GTIs from time bins where the count rate was below 3$\sigma$ of the mean, and applied this GTI file to the original events file.

While this method produced cleaned events files for most of the pseudo-exposures, it did a poor job in instances of intense flaring: a $3\sigma$-clipping was inadequate since the count rate distributions have an extended tail within the 3$\sigma$ tolerance level. For these pointings, we inspected the count rate distributions by eye to determine a cut-off value to remove the tail of this distribution, visually inspecting both sets of GTI-filtered events files to assess which filtering best removed the background to enhance signal from the sources.

Finally, we note that some pseudo-exposures were badly hampered by flaring such that no GTI filtering could recover useful signal. In Table \ref{xmm_summary}, we note which pseudo-exposures were subsequently discarded from our analysis, and whether this affected just the PN detector or all three detectors. We also indicate the effective area covered by each observation after removing flared pointings.

\subsection{Generating Products for Source Detection}
We extracted images from the GTI-filtered events files, using all valid events (PATTERN 0 to 12) for MOS1 and MOS2 and single to double events (PATTERN 0 to 4) for PN. We excluded the energy range from 1.45-1.54 keV to avoid the Al K$\alpha$ line (1.48 keV) from the detector background. The PN detector also has background emission lines from Cu at $\sim$7.4 keV and $\sim$8.0 keV, so we excluded the energy ranges from 7.2-7.6 keV and 7.8-8.2 keV when extracting PN images. To correct for the out-of-time events, the PN OoT images were scaled by 0.063 and subtracted from the PN images. We then extracted MOS and PN images in the standard soft (0.5-2 keV), hard (2-10 keV), and full (0.5-10 keV) energy ranges and coadded the images among the detectors.

Exposure maps, which quantify the effective exposure time at each pixel in the detector, accounting for vignetting, were generated with the SAS task {\it eexpmap} for each detector and energy range. Since vignetting is a strong function of energy, we spectrally weighted the exposure maps such that the mean effective energy inputted into {\it eexpmap} is determined by assuming a powerlaw model where $\Gamma$=2.0 in the soft band and $\Gamma$=1.7 for the hard and full bands \citep[see][]{cappelluti}. This spectral model was also used to calculate energy conversion factors (ECFs) to convert from count rates to flux, as summarized in Table \ref{ecf} \citep[for a discussion of how different assumptions for $\Gamma$ affect the derived ECF, see][]{loaring,cappelluti,ranalli}. The exposure maps were coadded among the detectors, weighted by their ECFs.

\begin{deluxetable}{lrr}
\tablewidth{225pt}
\tablecaption{\label{ecf} Energy Conversion Factors (ECFs)\tablenotemark{1}}
\tablehead{\colhead{Band} & \colhead{PN} & \colhead{MOS} } 

\startdata
Soft (0.5-2 keV)  & 7.45 & 2.00 \\
Hard (2-10 keV)  & 1.22 & 0.45 \\
Full (0.5-10 keV) & 3.26 & 0.97 
\enddata
\tablenotetext{1}{ECFs in units of counts s$^{-1}$/10$^{-11}$ erg cm$^{-2}$ s$^{-1}$. These are based on a spectral model where $N_{\rm H}=3\times10^{20}$ cm$^{-2}$ and $\Gamma$=2.0 in the soft band and $\Gamma$=1.7 in the hard and full bands. The PN ECF takes into account energy ranges that were masked out due to detector background line emission.}
\end{deluxetable}

As described in detail by \citet{me2}, we used the algorithm presented in \citet{cappelluti} to create background maps. In brief, a simple source detection was run on each detector image in each energy band using the SAS task {\it eboxdetect} with a low detection probability (likemin = 4). The positions of these sources were masked out. The remaining emission results from unresolved cosmic X-ray sources and local particle and detector background.  These components were modeled and fit as discussed in \citet{cappelluti} and \citet{me2} to produce a background map for each detector and energy range. The resulting background maps were then coadded among the detectors.

Before importing these products into the source detection software, we updated the header keywords ``RA\_NOM'', ``DEC\_NOM'', ``EXP\_ID'', and ``INSTRUME'' to common values among the pseudo-exposures for each observation: the SAS source detection software, when running on these files simultaneously, will fail if the pseudo-exposures do not have common WCS, exposure ID, and instrument values. However, the ``RA\_PNT'' and ``DEC\_PNT'' header keywords were manually updated to reflect the central coordinates of each pseudo-exposure so that the point spread function (PSF) is correctly calculated during source detection. Detector masks were created using the SAS program {\it emask}, which uses the exposure map as input to determine which pixels are active for source detection.

\subsection{Source Detection}
We produced a preliminary list of sources by running the SAS task {\it eboxdetect}  in ``map'' mode. This is a sliding-box algorithm that is run on the coadded images, background maps, exposure maps, and detector masks, where source counts are detected in a $5\times5$ pixel box with a low-probability threshold (likemin=4). This list is then used as an input into {\it emldetect} which performs a maximum likelihood PSF fit to the {\it eboxdetect} sources. We used a minimum likelihood threshold ($det\_ml$) of 6, where $det\_ml$= -ln$P_{\rm random}$, where $P_{\rm random}$ is the Poissonian probability that a detection is due to random fluctuations. We also included a fit to mildly extended sources, where {\it emldetect} convolves the PSF with a $\beta$ model profile. We consider a source extended if the output {\it ext\_flag} exceeds 0. Finally, the ECFs reported in Table \ref{ecf} were summed among the detectors included in the coadded pseudo-exposures (i.e., if the PN image was discarded due to flaring, the ECF sum is from the MOS detectors, while the PN ECF is included in the sum when all detectors are useable), such that {\it emldetect} reports the flux in physical units, as well as the count rates, for each detected source.

We ran the source detection algorithm separately for each observation. Due to the limited memory capabilities of the SAS source detection software, not all pseudo-exposures within an observation could be fit simultaneously. We therefore executed the source detection in batches, where adjacent rows in RA were fit simultaneously. To achieve the greatest coadded depth in the overlapping pointings, each column, other than the Eastern and Western edges of the mosaic, was included in two source detection runs. We note that the deepest overlap regions are fitted with this source-detection method. The source detection was also run separately for the different energy bands: soft (0.5-2 keV), hard (2-10 keV), and full (0.5-10 keV).

\subsection{Source List Generation}
From the above procedure, we have six source lists per energy band per observation. Each list contains duplicate detections of some sources due to the overlapping regions covered in consecutive source detection runs. To produce a clean X-ray source list for each observation, we removed these duplicate detections. Following the algorithm used by the {\it XMM-Newton} Serendipitous Source Catalog \citep{watson} to flag duplicate observations, we consider objects from source lists covering overlapping areas to be the same if the distance between them is less than $d_{\rm cutoff}$ where $d_{\rm cutoff}$ = min($0.9 \times d_{\rm nn1},0.9 \times d_{\rm nn2},15^{\prime\prime},3 \times (\sqrt{{\rm ra\_dec\_err^{2}_{1} + sys\_err^{2}}} + \sqrt{{\rm ra\_dec\_err^{2}_{2} + sys\_err^{2}}})$), where $d_{\rm nn1}$ ($d_{\rm nn2}$) is the distance between the source and its nearest neighbor in the first (second) source list, ra\_dec\_err is the X-ray positional error reported by {\it emldetect}, and sys\_err is a systematic positional error, taken to be 1$^{\prime\prime}$, to account for the sources not having an external astrometric correction applied. The maximum search radius of 15$^{\prime\prime}$ was chosen as the maximum cut-off distance based on simulations discussed in \citet{me2}, where we found that this radius maximizes identification of output to input sources while minimizing spurious associations of distinct sources; due to the shallow nature of our observations, source confusion from a high density of resolved sources is not a concern (see Section \ref{s82x_prop} for estimated source confusion rate). For duplicate detections of the same source, we retain the coordinate, flux, and count information for the object that has the highest detection probability, or $det\_ml$. We perform this routine separately for each energy band, producing one clean source list per band.

We then merge these X-ray source lists for each energy band of an observation using the search criterion defined above to find matches among lists generated in the separate energy bands. If no match is found, the source is considered undetected in that band and its flux, flux error, counts, and $det\_ml$ are set to null while we retain this information for the band(s) where it is detected. While we have discarded sources that are extended in all bands in which they are detected, because the identification of clusters
among the extended sources is in progress and will be reported later, we have flagged the sources that are point-like in one band and are extended in another band. The ``ext\_flag'' is non-zero for these objects and is defined as follows: 1 - extended in the soft band, 2 - extended in the full band, 3 - extended in the hard band, 4 - extended in the soft and full bands, 5 -  extended in the soft and hard bands, 6 - extended in the hard and full bands. 

To produce the final catalog, the coordinates are averaged among the coordinates from the individual energy band catalogs and their positional errors are added in quadrature; we note that the significance of the detection is not taken into account when averaging the coordinates, but the uncertainty in the astrometric precision is included by adding the positional errors in quadrature. We then retain only objects where $det\_ml$ exceeds 15 (i.e., $>5\sigma$) in at least one energy band \citep[see][for a discussion of $det\_ml$ limits and their effects on Eddington bias in the derived Log$N$-Log$S$ relation]{loaring, mateos}. 

We caution that care must be taken when determining the reliability of the reported fluxes as the catalog includes the {\it emldetect} reported fluxes for every band where the source was detected (i.e., $det\_ml \geq 6$). Though the X-ray source can be considered a significant detection as $det\_ml$ has to exceed 15 in at least one energy band for the source to be included in the catalog, the $det\_ml$ value for each band ought to be used to determine whether the reported flux is at an acceptable significance level. For reference, we use only fluxes in the subsequent analysis when $det\_ml \geq 15$ in that band.

Finally, we assign each X-ray source a unique record number (``rec\_no''), ranging from 2359 to 5220, since the previous {\it XMM-Newton} Stripe 82X catalog release terminated at ``rec\_no'' 2358. We also include columns ``in\_chandra'' and ``in\_xmm'' to note whether a source was detected in the archival {\it Chandra} or {\it XMM-Newton} Stripe 82X catalogs, respectively, as well as the corresponding identification number of the matched source; for the one {\it XMM-Newton} source that has two possible {\it Chandra} counterparts within the search radius (rec\_no 3473), due to {\it Chandra}'s superior spatial resolution, we list both of the {\it Chandra} matches. Details about each column are summarized in the Appendix.

\section{Stripe 82X Survey Sensitivity and Log$N$-Log$S$}\label{s82x_prop}
Similar to our previous Stripe 82X release, we gauge survey sensitivity for our {\it XMM-Newton} AO13 program via Monte Carlo simulations. For each observation, we generated a list of fluxes that follow published Log$N$-Log$S$ relations from {\it XMM}-COSMOS \citep{cappelluti2009} for the soft and hard bands and from ChaMP \citep{kim2007} for the full band. The minimum flux was set to 0.5 dex below the lowest detected flux in the source list for that observation and the maximum flux was set to 10$^{-11}$ erg s$^{-1}$ cm$^{-2}$. An input source list is then generated by pulling random fluxes from this distribution which are then given random positions among the pseudo-exposures making up the observation. We then use part of the simulator written for the {\it XMM-Newton} survey of CDFS \citep{ranalli} to convolve the input source list with the {\it XMM-Newton} PSF to create mock events files from which images were extracted. The observed background is then added to the simulated images. Since the exposure maps from the observations were used when generating the simulated events files and the observed background was added to the mock image, the simulations allow us to accurately gauge how well we can recover input sources given our observing conditions. Finally, we add Poissonian noise to the images and run these products through the source detection algorithm detailed above, using ancillary products (i.e., background maps, exposure maps, and detector masks) from the observations. We ran a suite of 20 simulations for each mosaicked observation.

Since we have both the input source list and the list of detected objects, we can estimate the spurious detection rate for our sample. We assume that any source detected above our $det\_ml$ threshold of 15 that does not have an input source within 15$^{\prime\prime}$ is a spurious detection. We find our spurious detection rate for the {\it XMM-Newton} AO13 data to be 1.0\%, 0.67\%, and 0.33\% in the soft, hard, and full bands, respectively. Furthermore, we can estimate the confusion fraction, which is when input sources are unresolved in the source detection and observed as one object. As we did in \citet{me2}, we followed the prescription in \citet{cappelluti} to test for source confusion, using the criterion $S_{\rm out}$/($S_{\rm in} + 3\sigma_{\rm out}$) $>$1.5, where $S_{\rm out}$ is the output flux, $S_{\rm in}$ is the input flux, and $\sigma_{\rm out}$ is the {\it emldetect}-reported flux error. According to this metric, the source confusion rate is 0.15\%, 0.10\%, and 0.16\% percent in the soft, hard, and full bands, respectively.

\begin{figure}
\centering
\includegraphics[angle=90,scale=0.4]{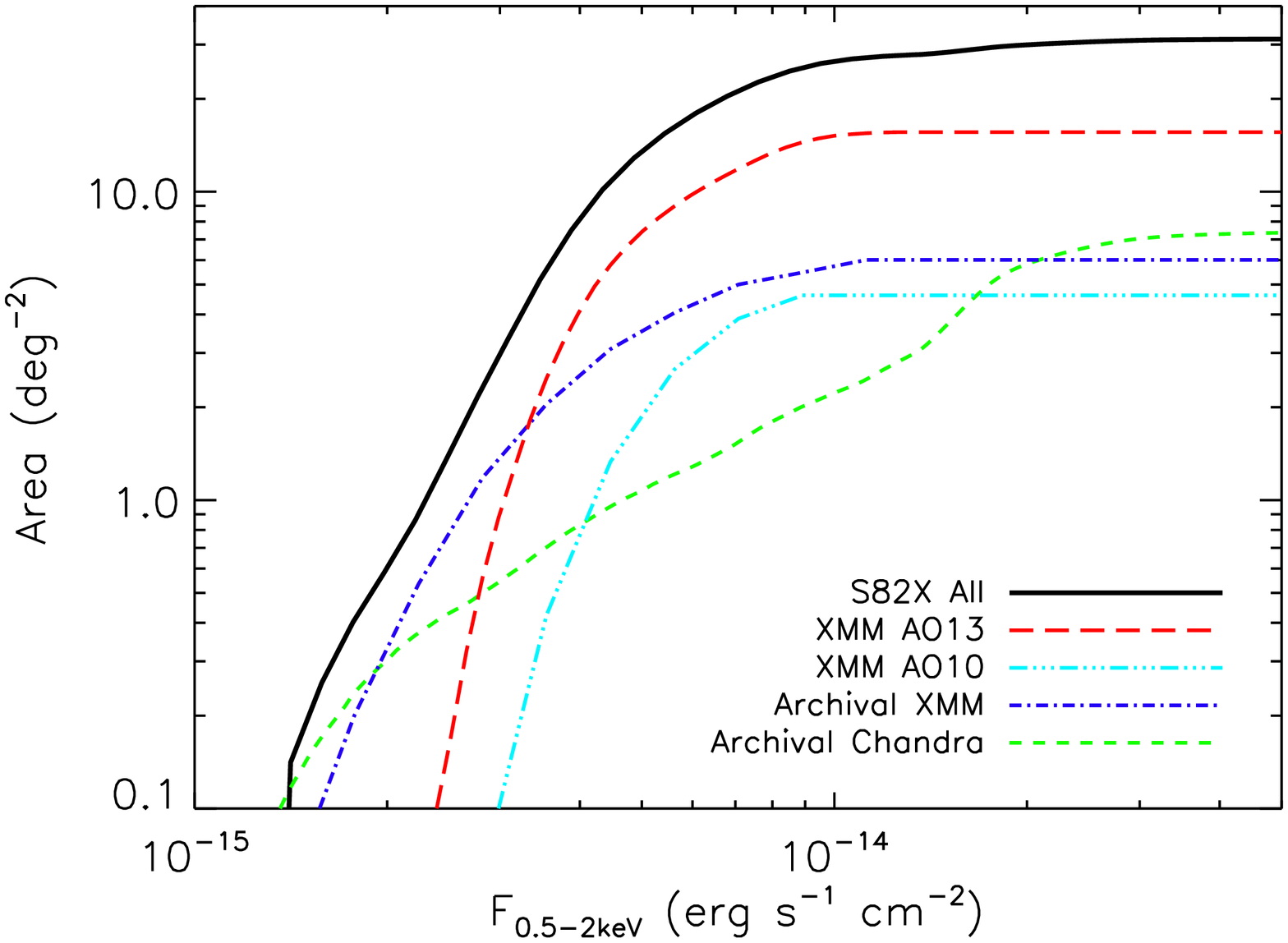}
\includegraphics[angle=90,scale=0.4]{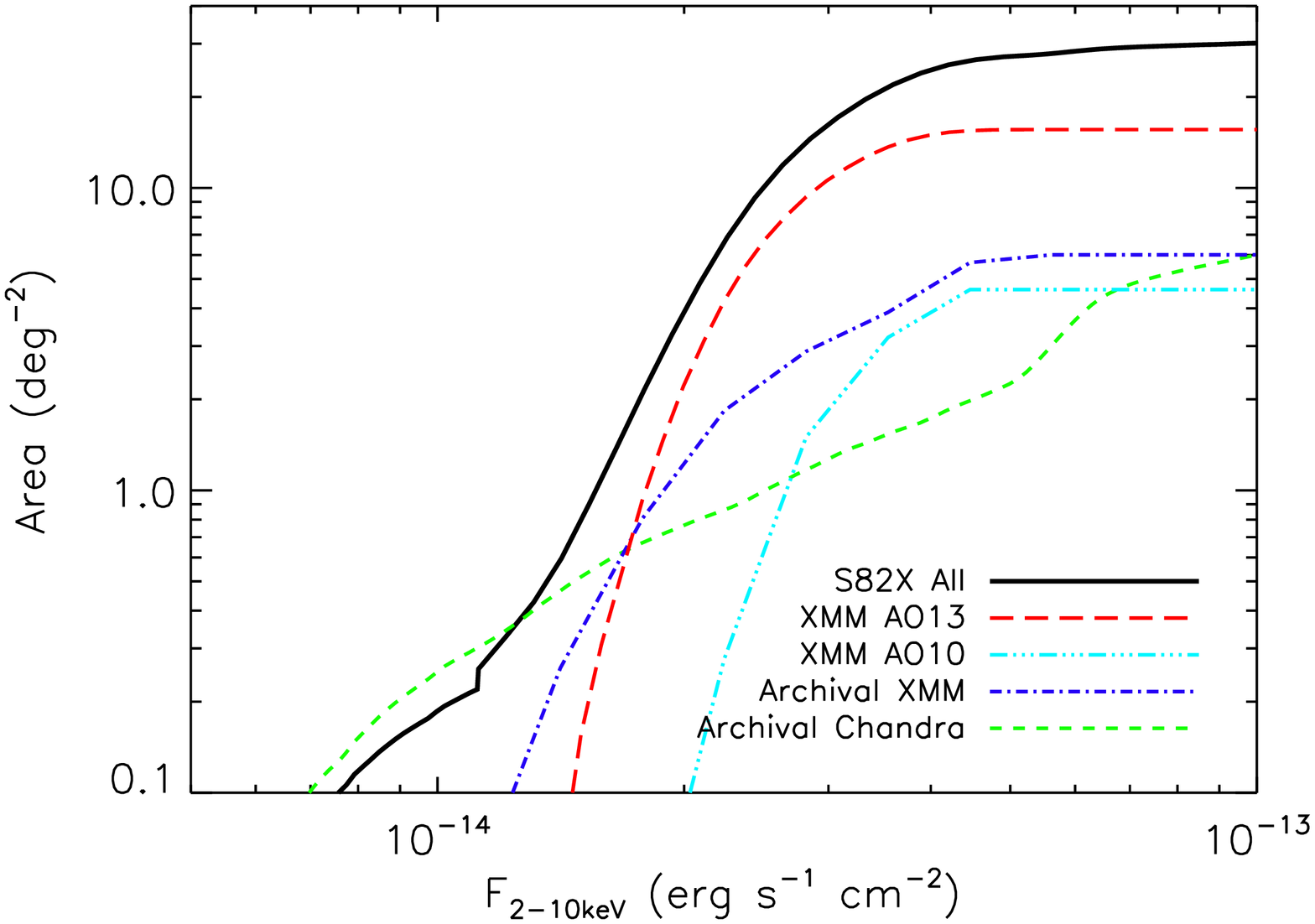}
\includegraphics[angle=90,scale=0.4]{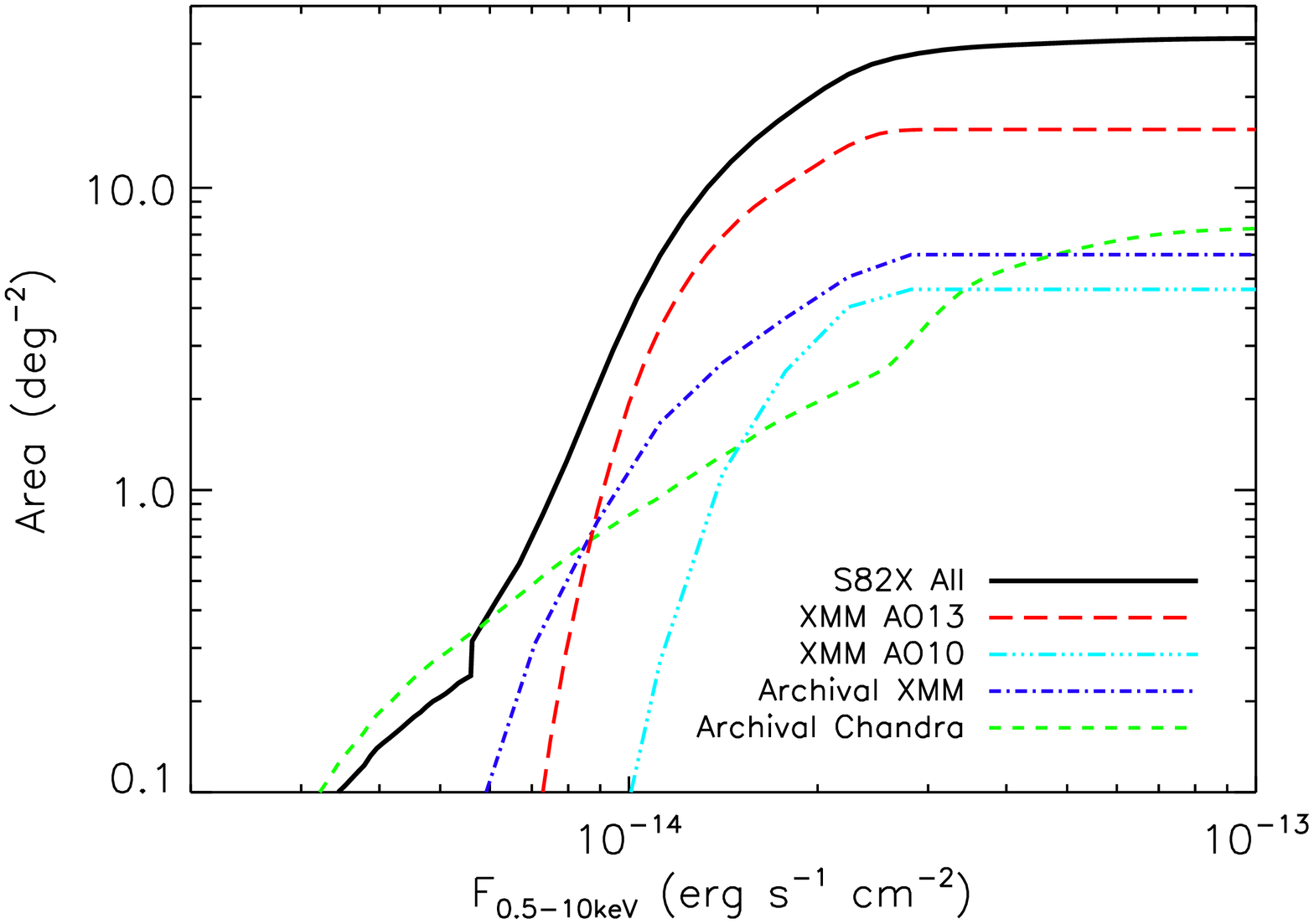}
\caption{\label{area_flux} Area-flux curves for Stripe 82X in the
 soft ({\it top}), hard  ({\it middle}), and full
  bands ({\it bottom}). While the colored curves show the full area for the
  individual datasets as indicated in the legends, the black curve
  illustrates the total area after removing observations from the
  archival {\it Chandra} and archival {\it XMM-Newton} surveys that
  overlap pointings from the {\it XMM-Newton} AO10 and/or AO13 surveys, and, in the
  case of the archival {\it Chandra} observations, archival {\it
    XMM-Newton} surveys; here, we given preference to the wider-area coverage from {\it XMM-Newton} in overlapping pointings. Hence, deeper fluxes accessible by {\it Chandra} are consequently removed from the total Stripe 82X area-flux relation. The kink in the total area-flux curve in the hard and full bands comes from discontinuties induced by combining the individual area-flux curves from the archival pointings at lower flux limits.}
\end{figure}

To determine survey sensitivity, we generate histograms of all input fluxes and output fluxes for the $det\_ml \geq 15$ sources, and divide the latter by the former. We truncate this ratio where it reaches unity. By multiplying this sensitivity curve, which is a function of flux, by the survey area, we derive the area-flux curves shown in red in Figure \ref{area_flux}. For comparison, we also plot the area-flux curves for the other components of the Stripe 82X survey in Figure \ref{area_flux}: archival {\it Chandra} (green), archival {\it XMM-Newton} (dark blue), and {\it XMM-Newton} AO10 (cyan). The black curve shows the total Stripe 82X area-flux relation after removing overlapping observations between the {\it Chandra} and {\it XMM-Newton} surveys and between the {\it XMM-Newton} archival and AO13 surveys. To convert the observed 2-7 keV and 0.5-7 keV {\it Chandra} bands to the {\it XMM-Newton}-defined hard and full bands of 2-10 keV and 0.5-10 keV, we used the assumed power-law model of $\Gamma=1.7$ \citep[see][]{me2} to extrapolate the {\it Chandra} flux to the broader energy ranges (i.e., the hard and full fluxes were multiplied by factors of 1.36 and 1.21, respectively).

In Table \ref{xray_summary}, we summarize the number of X-ray sources detected at a significant level for each Stripe 82X survey component. For the {\it XMM-Newton} surveys, a source is deemed significant if $det\_ml$ exceeds 15 in the specific energy band while for the {\it Chandra} survey, significance is determined by comparing the source flux at the pixel where it was detected with the 4.5$\sigma$ sensitivity map value at that pixel \citep[see][for details]{me1}. The ``Total'' row in Table \ref{xray_summary} removes duplicate observations of the same source in overlapping pointings among the survey components. In the current 31.3 deg$^2$ Stripe 82X survey, 6181 distinct sources are significantly detected between {\it XMM-Newton} and {\it Chandra}.

\begin{figure*}
\centering
\includegraphics[angle=90,scale=0.35]{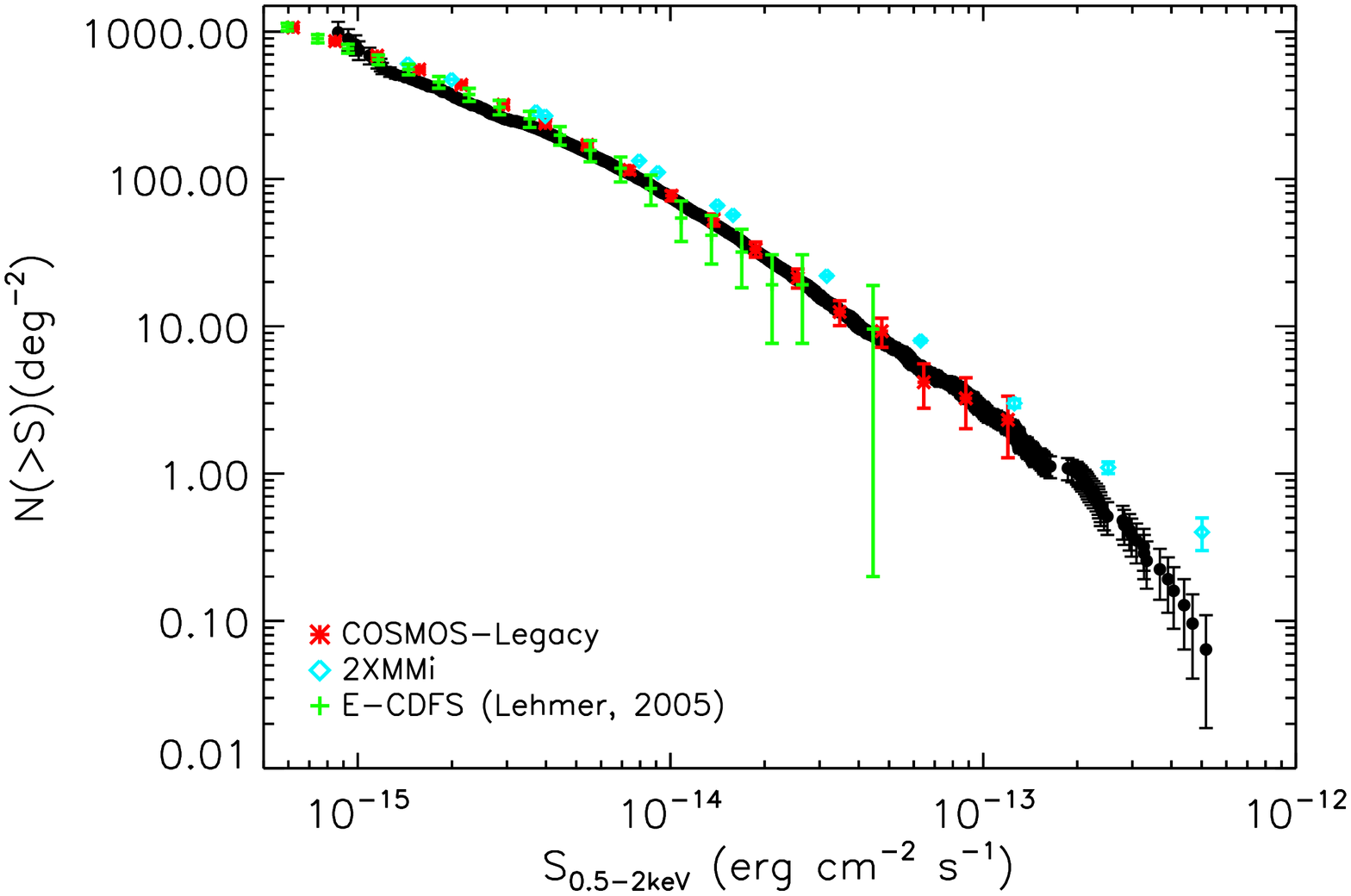}
\includegraphics[angle=90,scale=0.35]{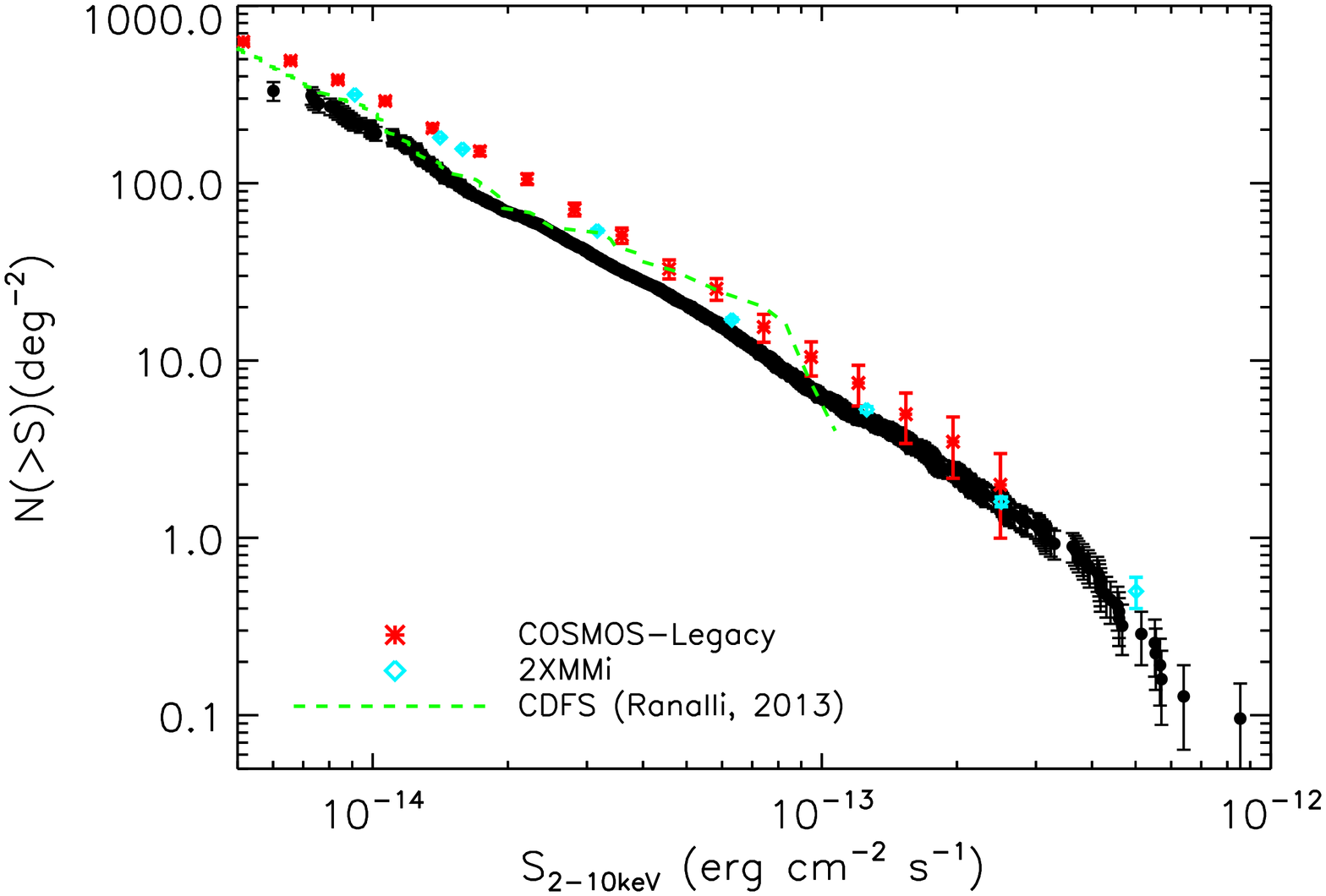}
\includegraphics[angle=90,scale=0.35]{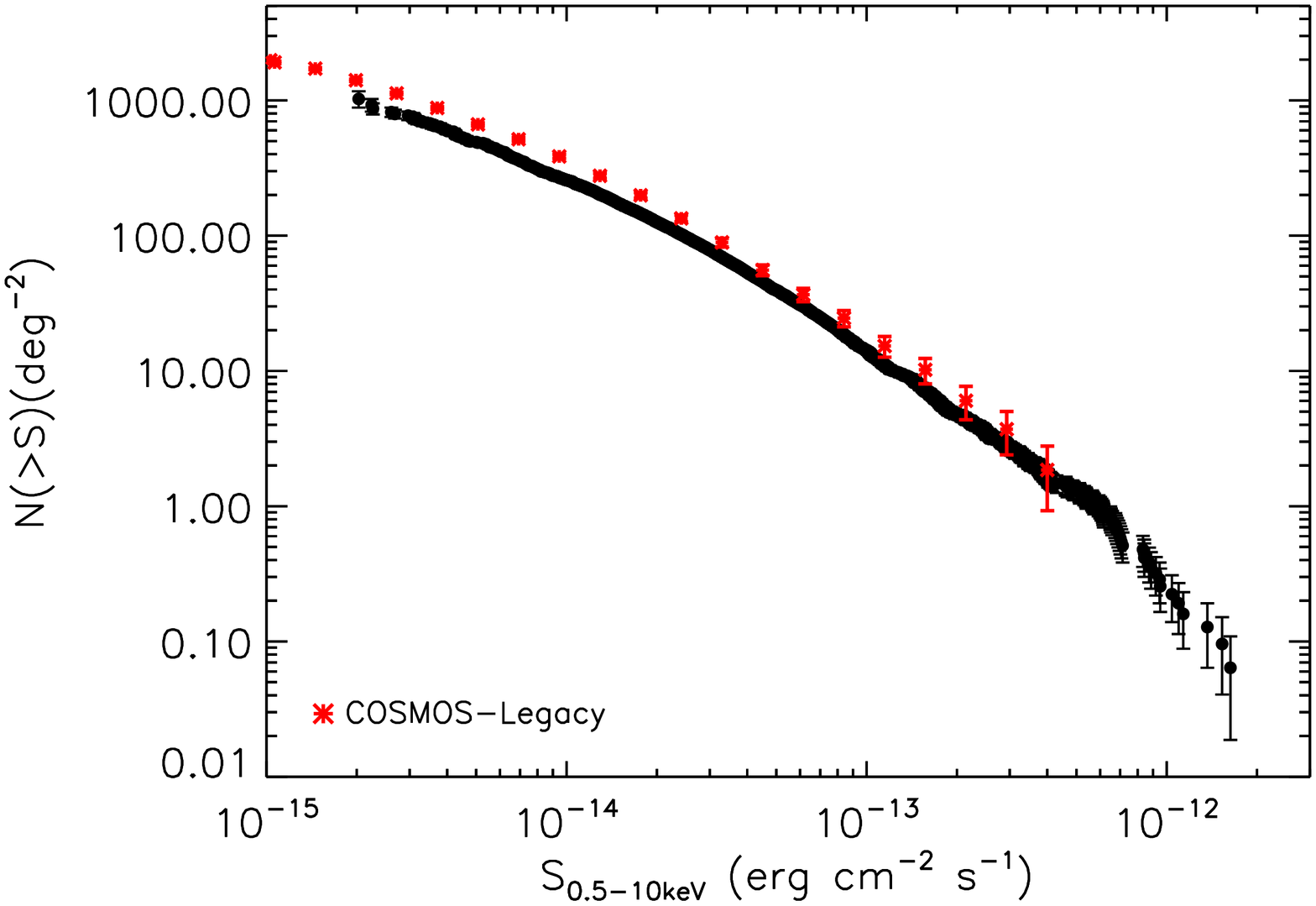}
\caption{\label{logn_logs} Cumulative Log$N$-Log$S$ relationship for the Stripe 82X survey ({\it black circles}) in the soft ({\it top left}), hard ({\it top right}) , and full ({\it bottom})  bands. For reference, we also plot the source number density for other surveys, spanning the gamut from deep, pencil-beam surveys \citep[i.e., the 0.3 deg$^2$ ECDF-S and $\sim$0.25 deg$^2$ CDFS;][respectively]{lehmer,ranalli}, to a moderate-area, moderate depth survey (the 2.2 deg$^2$ {\it Chandra} COSMOS-Legacy; Civano et al., {\it submitted}, Marchesi et al., {\it submitted}), and a wide-area survey \citep[the 132 deg$^2$ 2XMMi Serendipitous Survey;][]{mateos}.}
\end{figure*}

We present the Log$N$-Log$S$ distribution, or number source density as a function of flux, of the current 31.3 deg$^2$ Stripe 82X survey in Figure \ref{logn_logs}. To be consistent with the area-flux curves, we combined the X-ray source lists from the archival {\it Chandra}, archival and AO10 {\it XMM-Newton}, and AO13 {\it XMM-Newton} catalogs, removing all sources from observations that were discarded from the area-flux relation due to overlapping area. Targeted objects from archival observations were also removed as discussed in \citet{me1,me2}. We also note that while the {\it Chandra} Log$N$-Log$S$ relation we published in \citet{me1} had the cluster fields removed {\it a priori}, we have made no such cut here since, as we mentioned in that work, we found that including or excluding such fields made no noticeable difference in the source density calculation. The {\it Chandra} hard and full band fluxes from the source list were converted from the 2-7 keV and 0.5-7 keV ranges to 2-10 keV and 0.5-10 keV bands as described above. For reference, we also plot the Log$N$-Log$S$ for a range of survey areas and depths: the deep, pencil-beam E-CDFS in the soft band \citep[0.3 deg$^2$;][]{lehmer} and the {\it XMM-Newton} survey of CDFS in the hard band \citep[$\sim$0.25 deg$^2$;][]{ranalli}; the moderate-area, moderate-depth {\it Chandra} COSMOS-Legacy Survey (2.2 deg$^2$; Civano et al., {\it submitted}, Marchesi et al., {\it submitted.}) in all three bands; and the wide-area 2XMMi Serendipitous Survey in the soft and hard bands \citep[132 deg$^2$;][]{mateos}. The Stripe 82X Log$N$-Log$S$ agrees with the reported trends from other surveys in the soft-band, the high-flux end in the hard and full bands, and with CDFS at the low-flux end ($<2\times10^{-14}$ erg s$^{-1}$) in the hard band; discrepancies in these bands at lower fluxes (and between CDFS and COSMOS-Legacy and 2XMMi in the hard band at low fluxes) may be due to different methods for estimating survey sensitivity when generating area-flux curves and different assumed values for the powerlaw slope ($\Gamma$) when converting count rate to flux, and is not necessarily atypical when comparing number counts from different surveys.

\begin{deluxetable}{lrrrr}
\tablewidth{0pt}
\tablecaption{\label{xray_summary} X-ray Source Summary\tablenotemark{1}}
\tablehead{\colhead{Survey} & \colhead{Soft} & \colhead{Hard\tablenotemark{2}} &
  \colhead{Full\tablenotemark{3}} & \colhead{Total} } 

\startdata
Archival {\it Chandra} (7.4 deg$^2$)         &   969 & 248 & 1137 & 1146 \\
Archival {\it XMM-Newton} (6.0 deg$^2$) & 1438 & 432 & 1411 & 1607 \\
{\it XMM-Newton} AO10 (4.6 deg$^2$)     &   635 & 175 &  668  & 751   \\
{\it XMM-Newton} AO13 (15.6 deg$^2$)   & 2440 & 715 & 2597 & 2862 \\
\hline\\
Total (31.3 deg$^2$)\tablenotemark{4}   & 5150 & 1520 & 5628 & 6181
\enddata
\tablenotetext{1}{The numbers correspond to the significant detections
  in each band. For {\it Chandra}, this is a 4.5$\sigma$ level based
  on comparing the flux with the sensitivity map \citep[see][for
  details]{me1} and for the {\it XMM-Newton} surveys, the $det\_ml$
  has to exceed 15.}
\tablenotetext{2}{The hard band spans 2-10 keV for the {\it
    XMM-Newton} surveys but corresponds to 2-7 keV for the {\it
    Chandra} survey.}
\tablenotetext{3}{The broad band is 0.5-10 keV for the {\it XMM-Newton}
  surveys but ranges from 0.5-7 keV for the {\it Chandra} survey.}
\tablenotetext{4}{Duplicate observations of the same source and
  overlapping observations between surveys removed in total numbers.}
\end{deluxetable}

\section{Multi-Wavelength Catalog Matching}\label{multi_wav}
We searched for counterparts to the {\it XMM-Newton} AO13 sources in publicly available multi-wavelength databases: SDSS, {\it WISE} \citep{wright}, UKIDSS \citep{hewett,lawrence,casali}, VHS \citep{mcmahon},{\it GALEX} \citep{morrissey}, FIRST, and the {\it Herschel} Survey of Stripe 82 \citep[HerS;][]{viero}. To determine whether a multi-wavelength association to an X-ray source represents the true astrophysical counterpart rather than a chance coincidence, we use the maximum likelihood estimator method \citep[MLE;][]{sutherland} to match between the X-ray source lists and the ancillary catalogs. MLE takes into account the distance between an X-ray source and ancillary objects within the search radius, the astrometric errors of the X-ray and ancillary sources, and the magnitude distribution of ancillary sources in the background to determine whether a potential multi-wavelength counterpart is more likely to be a background source or a true match. This method has been implemented in many X-ray surveys to identify reliable counterparts \citep[e.g.,][]{brusa2007,brusa2010,laird2009,cardamone,luo,civano,me2,marchesi}.

Ancillary objects within the search radius ($r_{\rm search}$), which is set at 7$^{\prime\prime}$ for the {\it XMM-Newton} AO13 sources \citep[see][]{brusa2010,me2}, are assigned a likelihood ratio ($LR$) which is the probability that the correct counterpart is found within $r_{\rm search}$ divided by the probability that a background ancillary source is there by chance:
\begin{equation}
LR = \frac{q(m)f(r)}{n(m)}.
\end{equation}
Here, $q(m)$ is the expected normalized magnitude distribution of counterparts within $r_{\rm search}$ which is estimated by subtracting the histogram of sources found within the search radius from the histogram of background objects, where each histogram is normalized by the relevant search areas; $f(r)$ is the probability distribution of the astrometric errors\footnote{$f(r)$ is modeled as a two-dimensional Gaussian distribution where the X-ray and ancillary positional errors are added in quadrature.}; and $n(m)$ is the normalized magnitude distribution of sources in the background. The background sources are taken as the objects found in an annulus around each X-ray source with an inner radius of 10$^{\prime\prime}$ and outer radius of 45$^{\prime\prime}$; thus, sources that are potential counterparts, i.e., within $r_{\rm search}$, are removed from the background estimation \citep{brusa2007}. We note that the positional error for the X-ray sources includes a 1$^{\prime\prime}$ systematic error added in quadrature to the {\it emldetect} reported positional error to account for the lack of an external astrometric correction. This systematic astrometric error was not included in the previous release of the Stripe 82X catalog, and we subsequently found that bright X-ray sources tended to have their positional errors under-estimated by {\it emldetect}, such that counterparts were missed by the matching algorithm even though visual inspection of the X-ray sources and ancillary objects revealed bright multi-wavelength objects that are likely true matches \citep[see also][]{brusa2010}. Adding the 1$^{\prime\prime}$ systematic error recovered these associations. Accordingly, the archival {\it XMM-Newton} and AO10 catalogs published previously have been updated here.

From $LR$, a reliability value is then calculated for every source:
\begin{equation}
R = \frac{LR}{\Sigma_i(LR)_i + (1 - Q)},
\end{equation}
where $Q$ is the ratio of the number of X-ray sources that have ancillary objects within the search radius divided by the total number of X-ray sources; the $LR$ sum is over every potential counterpart within the search radius of the X-ray source. This calculation is performed independently for every waveband to which we match the X-ray source list. We use $R$ as a way to distinguish between true counterparts and chance associations. For X-ray sources that have more than one possible association within $r_{\rm search}$, we retain the potential counterpart with the highest reliability. To determine the critical reliability threshold above which we consider an association the true counterpart ($R_{\rm crit}$), we follow the methodology in \citet{me2}: we produced a catalog where we shifted the X-ray positions by random amounts and matched the multi-wavelength catalogs to these randomized positions. The resulting reliability distribution then gives us an estimate of the number of contaminating spurious associations above $R_{\rm crit}$.  We pick our $R_{\rm crit}$ threshold by examining the reliability histograms of the ``true'' matches, i.e., the original X-ray catalog, and the ``spurious'' matches, i.e., the catalog with randomized positions, in bins of 0.05 to determine where the fraction of spurious matches is $\sim$10\%. That bin then becomes our threshold $R_{\rm crit}$ value. 

As always, matching the X-ray source lists to ancillary catalogs is a balancing act between minimizing contamination from unassociated sources and maximizing counterpart identification. It is unavoidable that some true counterparts will be missed and that spurious associations will be promoted as real matches. In Sections 4.2 - 4.7 below, we note the number of spurious matches, i.e., number of X-ray sources with randomized positions meeting the $R_{\rm crit}$ threshold, to the number of total matches from the original X-ray catalog above $R_{\rm crit}$ to provide an estimate of the counterpart contamination. We also show in Figures \ref{rsep_r} - \ref{rsep_hrs} the cumulative distribution of counterpart and spurious association fraction as a function of $r_{\rm sep}$, the distance between the X-ray and counterpart coordinates, for objects exceeding $R_{\rm crit}$. We remind the reader, however, that in addition to the separation between the sources, the astrometric error on both the X-ray and counterpart coordinates, the magnitude of the potential counterpart, and magnitude distribution of background sources all contribute to the calculated reliability value reported in the published catalogs.

As the X-ray sources represent a menagerie of astronomical objects (stars, galaxies, obscured AGN, and unobscured AGN) they will have a range of spectral energy distributions and thus not have the same relative strength among all the wave-bands in each ancillary catalog. For example, heavily obscured AGN are much brighter in the redder optical and infrared bands, and would have optical magnitudes in the bluer bands more consistent with background sources, or perhaps even be dropouts in these bands, while the converse is true for unobscured AGN. We therefore match the X-ray source list separately to each band in the multi-wavelength catalogs, determine $R_{\rm crit}$ independently for each passband, and then merge the individual lists where we report the maximum $R_{\rm crit}$ values among the matches for that catalog. The only exception to this procedure for the MLE matching is {\it WISE} since the $W1$ band is the most sensitive filter; all {\it WISE} sources in Stripe 82 have detections in the $W1$ band so we do not miss any objects by matching to $W1$ only. A high level summary of the multi-wavelength matches to the {\it XMM-Newton} AO 13 data is presented in the fifth column of Table \ref{multiwav_summary}.

\subsection{Cross-matches Between X-ray Catalogs}
\label{ao13_xmatch}
For the X-ray sources that are repeated among the individual catalogs (archival {\it Chandra}, archival and AO10 {\it XMM-Newton}, and AO13 {\it XMM-Newton} catalogs), we checked their multi-wavelength counterpart matches against each other. In most cases, these are consistent, but in some instances, a counterpart is not found for an X-ray source in one catalog yet is in another. This situation can arise due to differences in X-ray positions and positional errors between the individual sources lists, as well as the differences in the magnitude distribution of background sources. If a counterpart is found in one X-ray catalog and not another, we promote that counterpart as a match in the latter catalog. To keep track of such promoted matches, we have included the following flags: ``ch\_cp\_flag'', ``xmm\_archive\_cp\_flag'', and ``xmm\_ao13\_cp\_flag'' to indicate which counterparts were promoted into that catalog based on MLE matching from the archival {\it Chandra} catalog, archival and AO10 {\it XMM-Newton} catalog, and AO13 {\it XMM-Newton} catalog, respectively. If these fields are empty, then the independent MLE matching to the individual catalogs gave consistent results. Otherwise, the following numbers indicate which multi-wavelength counterpart is the promoted match: 1 - SDSS counterpart found but photometry rejected for failing quality control checks; 2 - SDSS;  3 - redshift; 4 - {\it WISE} counterpart found but rejected for failing quality control checks; 5 - {\it WISE}; 6 - UKIDSS; 7 - VHS; 8 - {\it GALEX}.\footnote{None of the UKIDSS or VHS matches were rejected for compromised photometry.}

While the number of matches quoted in the text below refer to the sources above $R_{\rm crit}$ in each catalog, the tally in the Table \ref{multiwav_summary} include the promoted counterparts found from cross-matching the catalogs. The remainder of this section pertains to the multi-wavelength catalog matching to the {\it XMM-Newton} AO13 source list, while updates to the previous released Stripe 82X catalogs are discussed in the Appendix.

\subsection{SDSS}
\begin{figure}
\centering
\includegraphics[angle=90,scale=0.35]{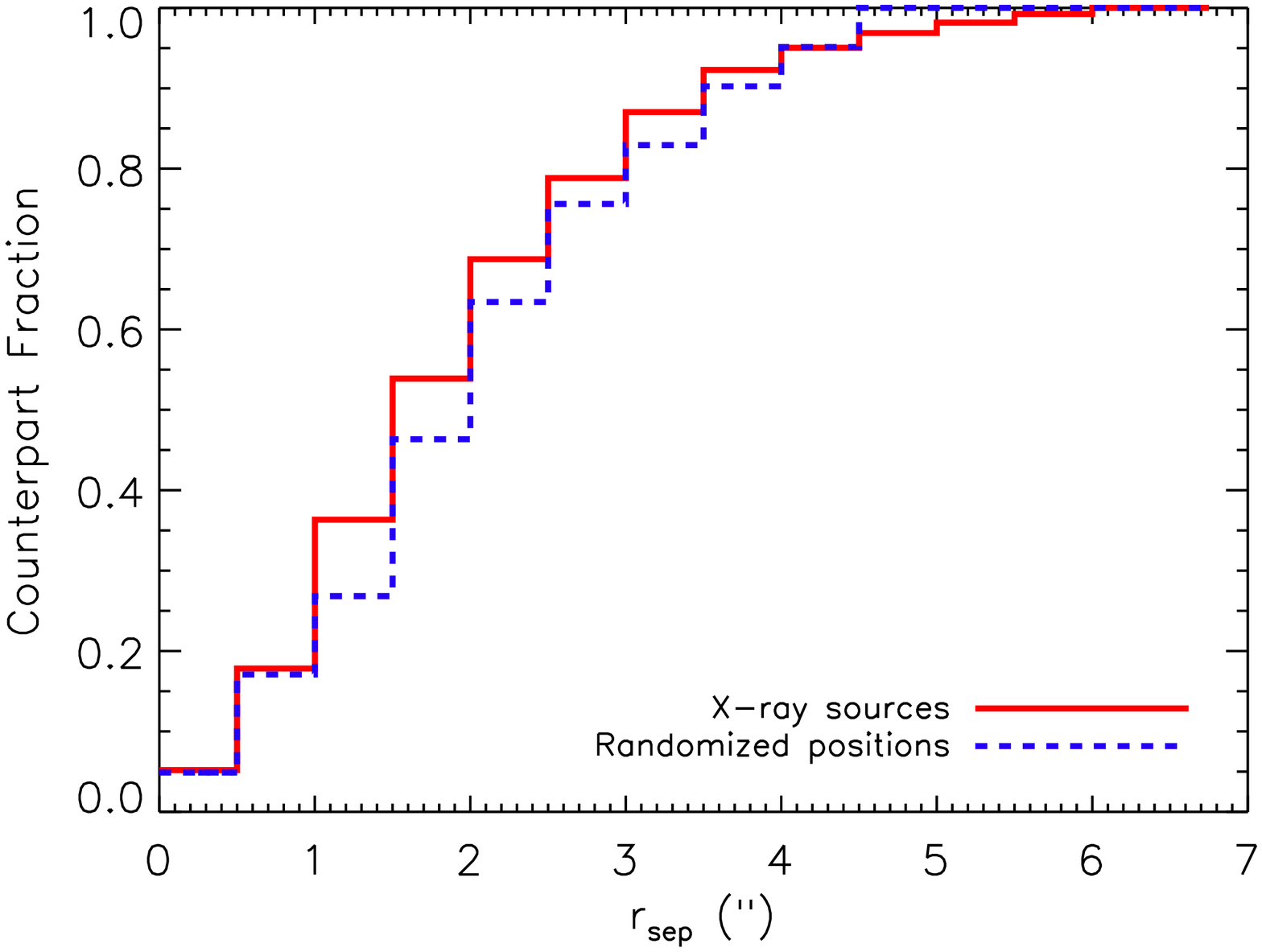}
\includegraphics[angle=90,scale=0.35]{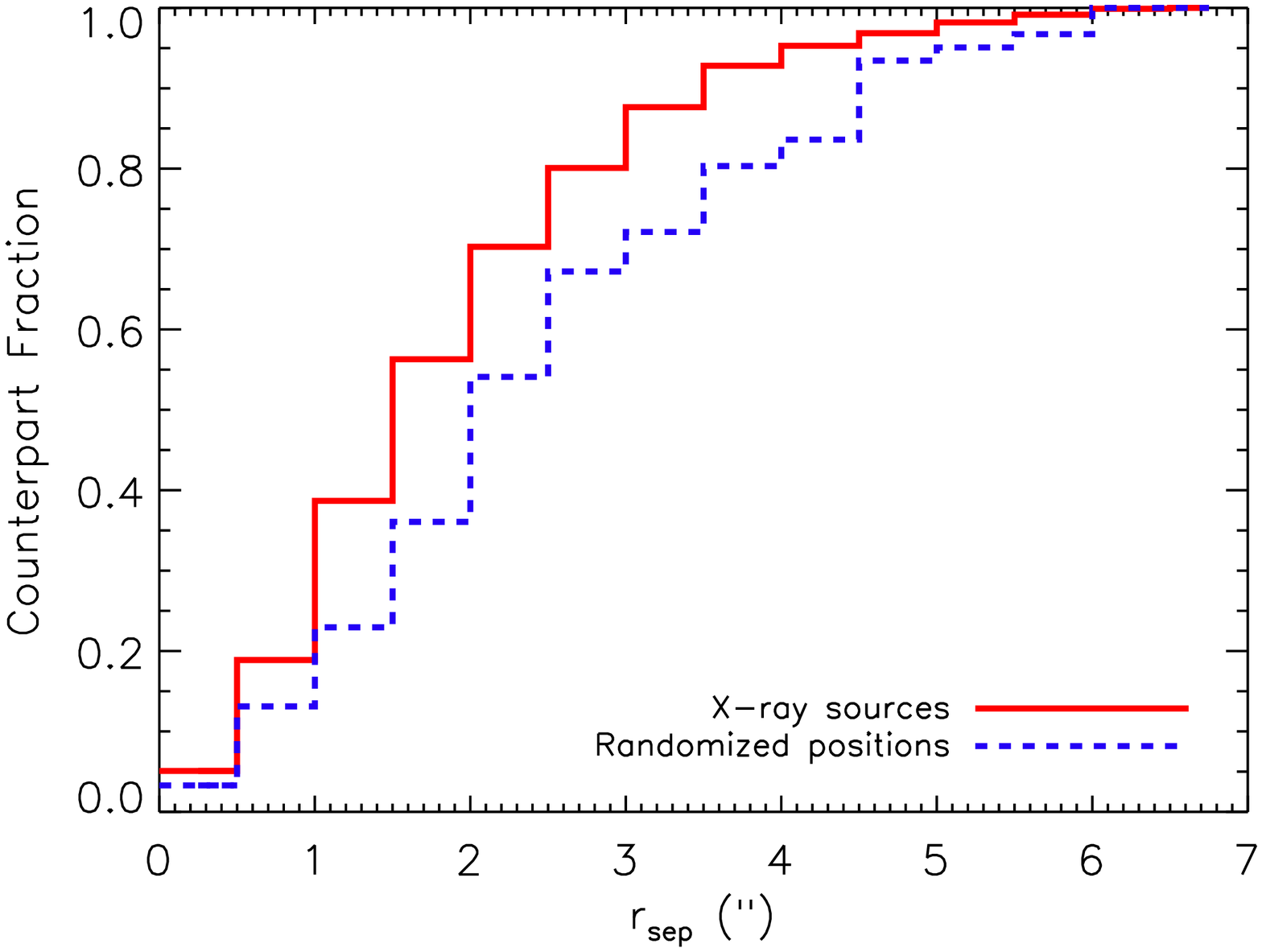}
\caption{\label{rsep_r} Cumulative distribution of the fraction of X-ray sources with an $r$-band counterpart above $R_{\rm crit}$ as a function of distance between the X-ray and SDSS positions ($r_{\rm sep}$; red solid line) and between the randomized X-ray positions and SDSS sources (blue dashed line). The {\it top} panels are for the matches from single-epoch imaging (1852 X-ray/SDSS counterparts and 41 random matches) while the {\it bottom} panels show the matches to the coadded \citet{jiang} catalog (1652 X-ray/coadded counterparts with 61 spurious associations). The number of spurious matches occurs at $r_{\rm sep}$ distances similar to that as the un-shifted X-ray catalog, indicating that MLE helps to mitigate unassociated sources compared to nearest neighbor matching by using magnitude and astrometric precision information in the calculation. }
\end{figure}

We matched the X-ray sources to the separate $u$, $g$, $r$, $i$, and $z$ bands in the single-epoch SDSS photometry from Data Release 9 \citep[DR9]{ahn}, where a uniform 0$\farcs$1 error was assumed for the SDSS astrometry \citep{rots}. We imposed the following $R_{\rm crit}$ values for the individual SDSS bands: $u$ - 0.75, $g$ - 0.80, $r$ - 0.85, $i$ - 0.85, $z$ - 0.80, with the estimated number of spurious association rate being 36/1989, 43/2006, 41/1852, 21/1819, 51/1926, respectively. Figure \ref{rsep_r} (top) shows the cumulative distribution of counterparts and spurious associations above the $r$-band $R_{\rm crit}$ value as a function of distance between the X-ray and SDSS source.

We removed from these individual band source lists any SDSS object that did not exceed the $R_{\rm crit}$ threshold, and then checked by eye the instances where more than one SDSS source is matched to an X-ray source to determine which optical source is the most likely counterpart. The preferred match is usually the SDSS source with the greatest number of matches among the individual bands and/or the brightest object. From our band-merged list, we then perform a photometric quality control to check for saturation, blending, or photometry that is not well measured.\footnote{We report the photometry for objects that meet the follow requirements: (NOT\_SATUR) OR (SATUR AND (NOT SATUR\_CENTER)), (NOT BLENDED) OR (NOT NODEBLEND), (NOT BRIGHT) AND (NOT DEBLEND\_TOO\_MANY\_PEAKS) AND (NOT PEAKCENTER) AND (NOT NOTCHECKED) AND (NOT NOPROFILE). An object that failed any of these quality control checks has the photometry set to -999 in the catalog.} Objects that do not meet these requirements are flagged in the ``SDSS\_rej'' column as ``yes'' in the catalog, though we retain the SDSS coordinates and ObjID to note that these sources are optically detected even if the photometry is compromised.  Finally, we check the remaining images by eye to remove optical artifacts, such as diffraction spikes and noise due to emission from nearby bright objects. 

We then matched the full X-ray catalog to the coadded SDSS source lists presented in \cite{jiang}, which are 1.9-2.2 mag deeper than the single-epoch SDSS imaging, with 5$\sigma$ magnitude limits of 23.9, 25.1, 24.6, 24.1, and 22.8 (AB) in the $u$, $g$, $r$, $i$, and $z$ bands. Here, we utilize the mag\_auto fields returned by SExtractor \citep{bertin} for the MLE algorithm. \citet{jiang} performed the image coaddition by separating each of the 12 SDSS parallel scans that cover Stripe 82 into 401 individual regions, extracting aperture magnitudes separately for each of the 5 bands. They then provide 24,060 individual catalogs, where each band, region, and scan line are independent catalogs, which can include duplicate observations of the same source among these catalogs that cover adjacent area. Thus, we first produced ``cleaned'' SDSS coadded catalogs by only retaining objects within 45$^{\prime\prime}$ of the {\it XMM-Newton} AO13 sources, since these are the data we need to estimate the background and find counterparts. We then search for duplicate observations within each band by searching for matches within 0$\farcs$5, retaining the coordinates and photometry for the object that has the highest signal-to-noise. We match the X-ray sources to each of these cleaned coadded catalogs. Here, the astrometric errors in the coadded images are similar to those of the single-epoch images due to the method used when generating the coadds \citep{jiang}. However, we conservatively used a value of 0$\farcs$2 based on observed positional offsets between SDSS coadded sources and FIRST objects (McGreer priv. comm.). We find the following $R_{\rm crit}$ cut-offs: $u$ - 0.85, $g$ - 0.9, $r$ - 0.9, $i$ - 0.9, $z$ - 0.85, with the spurious association rate being 20/1799, 41/1751, 61/1652, 40/1530, and 37/1816, respectively; the cumulative fraction of matches as a function of $r_{\rm search}$ above $R_{\rm crit}$ for both the X-ray source list and randomized positions is shown in the bottom panel of Figure \ref{rsep_r}. We note that the lower number of sources here compared with the single-epoch imaging data is due to the higher reliability thresholds we impose for the coadded catalog. However, the number of spurious associations in the lower reliability bins becomes a much higher fraction of the total number of true X-ray sources in those bins, so we have erred on the side of caution to minimize the number of random associations in our sample. 

From these counterparts found from matching to the coadded images, we keep only the sources that do not have a counterpart in the single-epoch imaging. Since \citet{jiang} do not provide a band-matched catalog or cross-identify the same source among the multiple-bands, we consider an optical source to be the same object if it is within $\sim$0$\farcs$6 with no other object found in that band within 1$^{\prime\prime}$; if no match in another band is found meeting these requirements, the source is assumed to be a drop-out in that band. The reported SDSS coordinates are the average of the coordinates in the individual band catalogs where the source is detected. The objects found from the coadded catalog are marked in the ``SDSS\_coadd'' column as ``yes''. 

In total, we find SDSS counterparts for 2438 X-ray sources (85\% of the sample), 178 of which are not found in the single-epoch SDSS imaging but are detected in the coadded catalog, and as expected are generally fainter. We list the information for the SDSS counterparts found from the single-epoch catalog, where available, to allow the user to easily query the main SDSS database to fetch relevant information using the unique SDSS ObjID or SDSS coordinates; similar data, such as aperture magnitudes and errors, from the coadded \citet{jiang} catalog would involve querying 24,060 individual catalogs, while such data are linked in the main SDSS database.

\subsubsection{Optical Spectra}
\label{spectra}
We mined the following public spectroscopic catalogs to find redshifts, and where possible, optical classifications of the X-ray sources with SDSS counterparts: SDSS Data Release 12 \citep[DR12;][]{dr12}, 2SLAQ \citep{croom}, pre-BOSS pilot survey using Hectospec on MMT \citep{ross_mmt}, and 6dF \citep{jones2004,jones2009}. We checked by eye the 41 spectra that had the {\it zwarning} flag set by the SDSS pipeline. While we were able to verify some of these redshifts, we were not able to find a reliable redshift solution for 26 of these objects, and set their redshifts to zero in the catalog. We also obtained spectra for 12 and 6 sources in 2014 September and 2015 January, respectively, through our dedicated follow-up program with WIYN HYDRA; the spectra were reduced with the {\it IRAF} task {\it dohydra} where we identified redshifts based on emission and/or absorption features, or classified stars on the basis of their rest-frame absorption and emission lines. About 29\% of the X-ray sources (828 objects) have secure redshifts. The calculation of photometric redshifts for the remainder of the sources is underway (Ananna et al. in prep.).

The databases we mined provide an automatic classification of sources based on their optical spectra, where ``QSO''s or ``AGN'' are objects that have at least one broad emission line in their spectra (generally a full-width half max exceeding 2000 km s$^{-1}$). Sources lacking broad emission lines are classified as ``galaxies,'' where this type includes objects with narrow emission lines \citep[Type 2 and elusive AGN, i.e. those objects with emission line ratios consistent with star-forming galaxies;][]{bpt,maiolino}, absorption lines only, and even blazars with featureless optical spectra that are not flagged as active galaxies by optical spectroscopic pipelines. We have followed this methodology when classifying sources from our spectroscopic follow-up campaign, where we reserve the class QSO to refer to broad-line objects and galaxies for sources lacking broad-lines. Stars are identified by emission and absorption transitions in their optical spectra.

\subsection{WISE}
Since publishing our initial Stripe 82X multi-wavelength matched catalogs in \citet{me2}, the AllWISE Source Catalog was released, combining data from the cryogenic and NEOWISE missions \citep{wright,mainzer}. As this catalog has enhanced sensitivity and astrometric precision, we match the {\it XMM-Newton} AO13 X-ray source list to this release, and update the archival {\it Chandra} and {\it XMM-Newton} and {\it XMM-Newton} AO10 matches to AllWISE, as detailed in the Appendix.

\begin{figure}
\centering
\includegraphics[angle=90,scale=0.35]{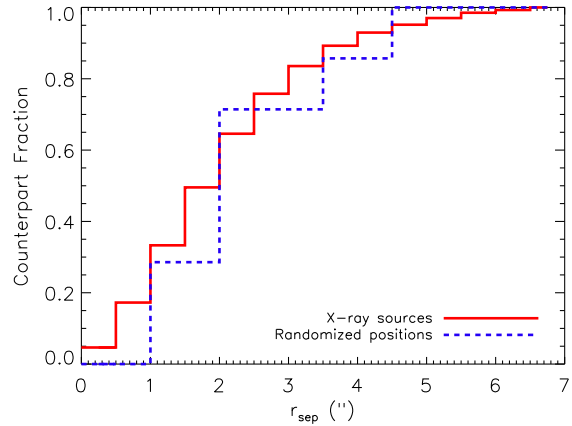}
\caption{\label{rsep_w1} Similar to Figure \ref{rsep_r}, but for the X-ray/{\it WISE} matches to the $W1$ band, where 2087 counterparts (7 spurious associations) are found (before discarding those failing quality control checks) above $R_{\rm crit}$. }
\end{figure}

When doing the MLE matching to the $W1$ band, using the ``w1mpro'' magnitude measured via profile-fitting photometry, we use a $R_{\rm crit}$ of 0.9, with 7 spurious associations out of 2087 matches (see Figure \ref{rsep_w1}). We then impose photometry control checks on the {\it WISE} sources, following our prescription in \citet{me2}. We null out the magnitude in any band that was saturated (i.e., the fraction of saturated pixels, ``w{\it n}sat'' exceeds 0.05, where $n$ refers to the band number); is likely a spurious detection associated with artifacts such as diffraction spikes, persistence, scattered light from nearby bright sources (i.e., if the ``cc\_flag'' is non-zero); or moon level contamination (i.e., if ``moon\_lev'' $\geq$ 5, where ``moon\_lev'' is the ratio of frames affected by scattered moonlight to the total number of frames and spans from 0 to 9). We also isolate extended sources as their ``w{\it n}mpro'' magnitudes would be unreliable. These sources have the ``ext\_flag'' set to non-zero. For these objects, we downloaded their elliptical photometry magnitudes (``w{\it n}gmag'') and discard their photometry if their extended photometry magnitude flags were non-null. If a matched {\it WISE} source has photometry that fails the point-like or extended photometry quality checks in all bands, then the ``wise\_rej'' flag is set to ``yes'' in the catalog and the associated photometry and coordinates are not reported.

Of the 2087 matched sources, 2031 (71\% of the {\it XMM-Newton} AO13 sources) passed the quality assurance tests above. All the rejected sources were extended. Ten extended sources had non-flagged elliptical magnitude measurements and are marked with the ``wise\_ext'' flag set to ``yes'' in the catalog.

\subsection{Near-Infrared}
The {\it XMM-Newton} AO13 source list was matched independently to the near-infrared (NIR) catalogs from the UKIDSS Large Area Survey \citep[LAS;][]{hewett,lawrence,casali,warren} and VHS \citep{mcmahon}. From both catalogs, we chose primary objects from the database\footnote{priOrSec = 0 OR priOrSec=frameSetId} and eliminated objects that were consistent with noise, i.e., ``mergedclass'' set to zero and ``pnoise''\footnote{``Pnoise'' is the probability that the detection is noise.} $>$0.05. The magnitudes presented in the catalog are the ``apermag3'' values from the UKIDSS LAS and VHS databases which are aperture-corrected magnitudes, with a 2$^{\prime\prime}$ diameter aperture.

\begin{figure}
\centering
\includegraphics[angle=90,scale=0.35]{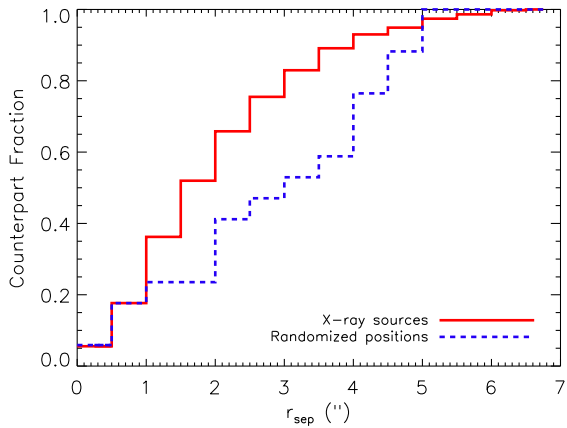}
\caption{\label{rsep_ukidss_k} Similar to Figure \ref{rsep_r}, but for the X-ray/UKIDSS matches to the $K$ band, where 1314 counterparts and 17 matches to randomized positions are found above $R_{\rm crit}$. }
\end{figure}

We matched the {\it XMM-Newton} AO13 source list to Data Release 8 of the UKIDSS LAS survey. Matching separately to the $Y$ (0.97 - 10.07 $\mu$m), $J$ (1.17 - 1.33 $\mu$m) , $H$ (1.49 - 1.78 $\mu$m), and $K$ (2.03 - 2.37 $\mu$m) bands, we find $R_{\rm crit}$ values of 0.75, 0.85, 0.75, 0.75, respectively, with a spurious association rate of 21/1375, 15/1070, 18/1335, and 17/1314, respectively (see Figure \ref{rsep_ukidss_k}). When merging the separate lists together, we find a total of 1784 near-infrared counterparts, or 62\% of the X-ray sample. We performed quality control checks on the photometry as explained in \citet{me2} to check for saturation, but no objects were flagged as being possibly saturated.

\begin{figure}
\centering
\includegraphics[angle=90,scale=0.35]{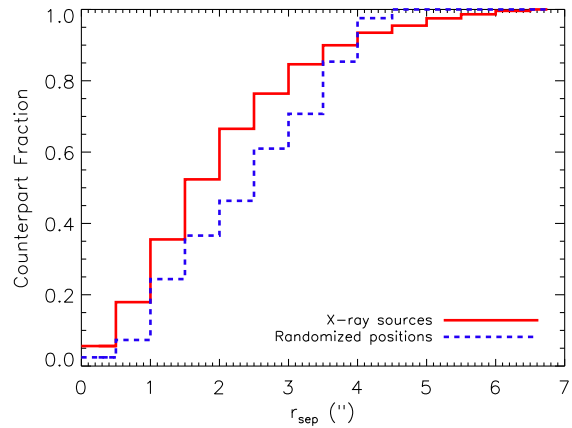}
\caption{\label{rsep_vhs_k} Similar to Figure \ref{rsep_r}, but for the X-ray/VHS matches to the $K$ band, where 1763 counterparts and 41 spurious associations are above $R_{\rm crit}$. }
\end{figure}

We used Data Release 3 of the VHS survey to match to the {\it XMM-Newton} AO13 catalog, where we adopt an astrometric uncertainty of 0$\farcs$14 for the VHS sources. VHS has coverage over Stripe 82 in the $J$, $H$, and $K$ bands, where we impose $R_{\rm crit}$ values of 0.75 in each band, with a spurious counterpart rate of 20/1856, 39/1783, and 41/1763, respectively (see Figure \ref{rsep_vhs_k}). In total, 2117 {\it XMM-Newton} AO13 sources (74\% of the sample) have NIR counterparts from the VHS survey. We also check the ``mergedClass'' flag to test if a source is saturated (``mergedClass''=-9), but none of the matches are so afflicted.

Between UKIDSS and VHS, we find NIR counterparts for 2257 X-ray sources, or 79\% of the sample. One hundred forty of the NIR sources are found in UKIDSS, but not VHS. Of these, 34 were non-detections in VHS (i.e., no match between VHS and UKIDSS within a 2$^{\prime\prime}$ search radius), while the remaining 106 were found in VHS but fell below our reliability thresholds for this catalog; we note that 77 of these VHS sources below the reliability cut had UKIDSS $Y$-band reliabilities above our $Y$-band critical threshold, while VHS is lacking this coverage. By presenting matches to both UKIDSS and VHS, the variability of the 1678 X-ray selected, NIR objects (1644 objects in common between UKIDSS and VHS and the 34 VHS dropouts) to be studied by the community.

\subsection{GALEX}
\begin{figure}
\centering
\includegraphics[angle=90,scale=0.35]{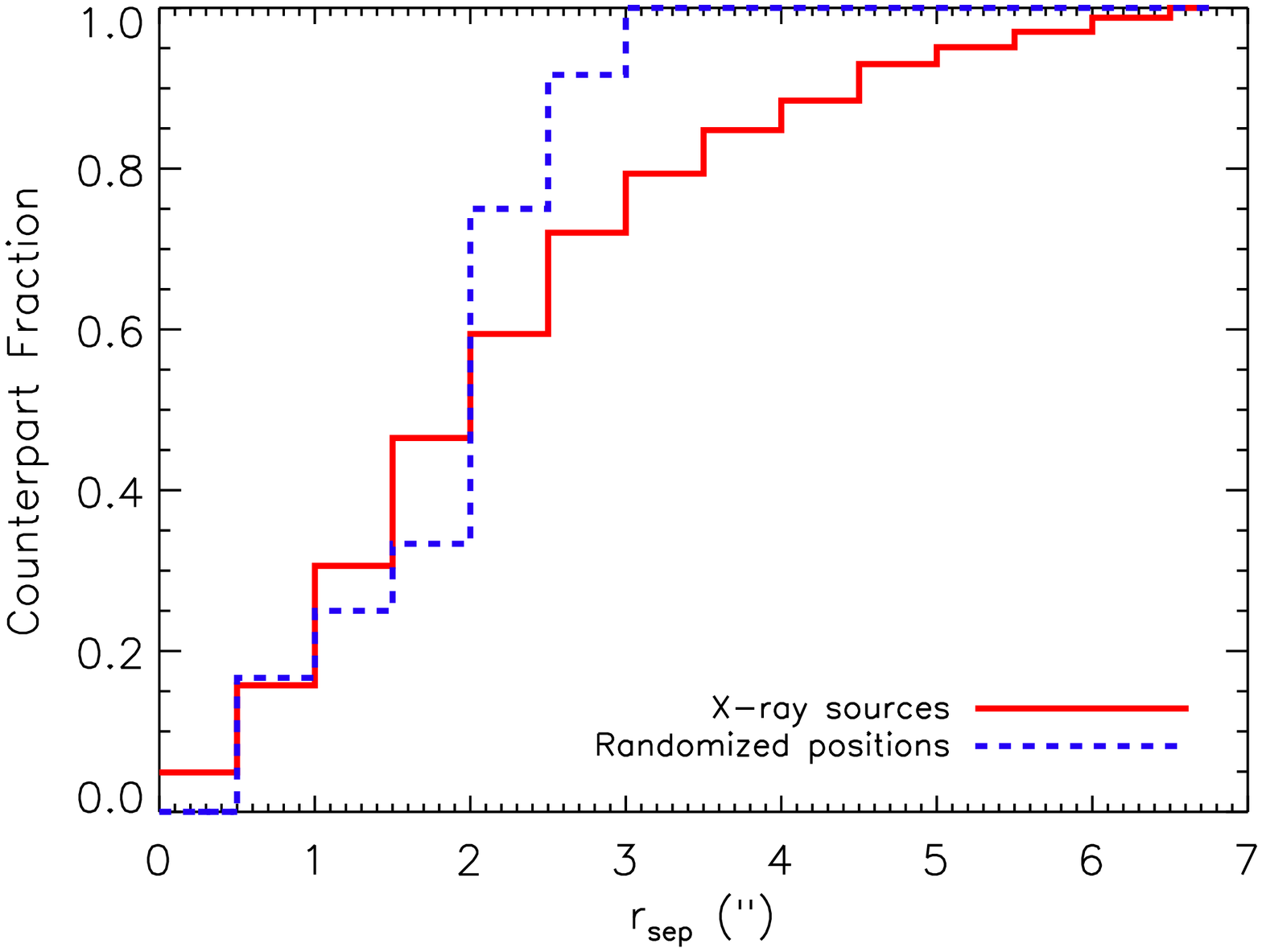}
\caption{\label{rsep_nuv} Similar to Figure \ref{rsep_r}, but for the X-ray/{\it GALEX} matches to the NUV band, where 572 counterpartsand 12 matches to randomized positions lie above $R_{\rm crit}$. }
\end{figure}

Similar to the UKIDSS matching, we used the cleaned {\it GALEX} catalog described in \citet{me2} to find counterparts to the {\it XMM-Newton} AO13 sources, matching to the near-ultraviolet (NUV) and far-ultraviolet (FUV) bands independently. This catalog represents data from the medium-imaging survey (MIS) in {\it GALEX} Release 7 \citep{morrissey}. With a $R_{\rm crit}$ value of 0.75 for both bands, we find 572 and 407 counterparts, with 12 and 5 spurious associations, in the NUV and FUV bands, respectively (see Figure \ref{rsep_nuv}). In total, 607 X-ray sources have ultraviolet counterparts, corresponding to 21\% of the {\it XMM-Newton} AO13 sample.

\subsection{FIRST}
\begin{figure}
\centering
\includegraphics[angle=90,scale=0.35]{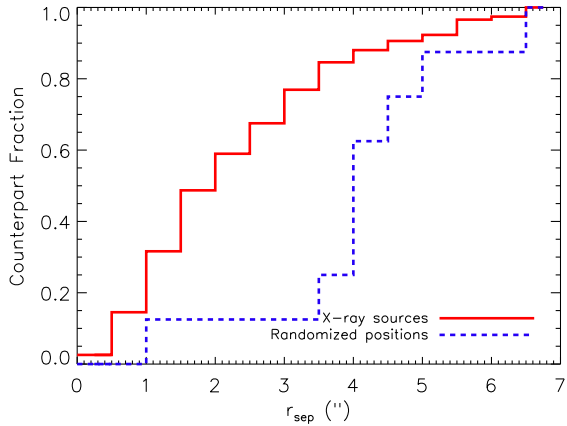}
\caption{\label{rsep_first} Similar to Figure \ref{rsep_r}, but for the X-ray and FIRST nearest-neighbor matches, with 116 counterparts and 8 randomized matches found within $r_{\rm search} = 7^{\prime\prime}$. Here, many of the spurious associations are found at higher separation distances due to the low number density of radio and X-ray sources.}
\end{figure}

Due to the relatively low space density of the radio sources detected in the Faint Images of the Radio Sky at Twenty centimeters \citep[FIRST;][]{becker1995,white} survey, we used a nearest neighbor match to find counterparts to the X-ray sources, using the same search radius of 7$^{\prime\prime}$ as employed in the MLE matching above. Similar to our previous Stripe 82X catalog release, we used the FIRST catalog published in 2012 which includes all sources detected between 1993 and 2011, which has a 0.75 mJy flux limit over the {\it XMM-Newton} AO13 region \citep{becker2012}. Since our previous paper, the final FIRST catalog has been published \citep{helfand} but we do not gain any additional sources when matching to this final catalog, both with the {\it XMM-Newton} AO13 data and archival {\it Chandra} and archival and AO10 {\it XMM-Newton} catalogs. One hundred sixteen FIRST sources (4\% of the X-ray sample) are found within 7$^{\prime\prime}$ of the {\it XMM-Newton} AO13 sources. When matching the FIRST catalog to the randomly shifted X-ray positions, 8 spurious associations were found (see Figure \ref{rsep_first}).

\subsection{Herschel}
The {\it Herschel} Stripe 82 Survey (HerS) covers 79 deg$^2$ at 250, 350, and 500 $\mu$m to an average depth of 13.0, 12.9, and 14.8 mJy beam$^{-1}$ at $>3\sigma$, surveyed with the Spectral and Photometric Imaging Receiver (SPIRE) instrument \citep{viero}. The far-infrared emission from {\it Herschel} provides a clean tracer of host galaxy star-formation \citep{pier}, making these data of particular importance to study the host galaxies of AGN \citep{pier,efstathiou,fritz,schartmann,lutz,schweitzer,netzer,shao,mullaney,rosario,magdis,delvecchio}. Indeed, the {\it XMM-Newton} AO13 survey was specifically designed to overlap existing {\it Herschel} coverage, since similar far-infrared data will not be available in the foreseeable future.

\begin{figure}
\centering
\includegraphics[angle=90,scale=0.35]{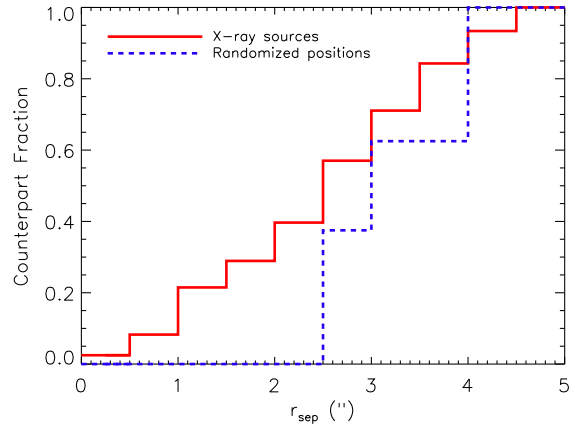}
\caption{\label{rsep_hrs} Similar to Figure \ref{rsep_r}, but for the X-ray and {\it Herschel} nearest-neighbor matches, with 121 counterparts and 8 spurious associations within $r_{\rm search} = 5^{\prime\prime}$. Like Figure \ref{rsep_first} shows with the matching between the X-ray source list and FIRST, most of the spurious matches occurs at higher values of $r_{\rm sep}$.}
\end{figure}

Similar to the matching to the FIRST catalog, we employed a nearest neighbor approach to find associations between the far-infrared {\it Herschel} sources and the X-ray objects. However, we shortened $r_{\rm search}$ to 5$^{\prime\prime}$ since our exercise of matching the {\it Herschel} catalog to the random X-ray positions reveals that most spurious associations occurred at distances between $5^{\prime\prime} - 7^{\prime\prime}$. We found 121 {\it Herschel} sources within 5$^{\prime\prime}$ of the {\it XMM-Newton} sources, corresponding to 4\% of the sample, and 8 spurious associations when matching to the randomized X-ray positions (see Figure \ref{rsep_hrs}).

\subsection{{\it XMM-Newton} AO13 Multiwavelength Match Summary}

In total, we find counterparts to 93\% of the {\it XMM-Newton} AO13 sources. However, we emphasize that we matched the X-ray source list independently to each of the multi-wavelength catalogs and {\it did not cross-correlate the counterparts.} In a vast majority of the cases, these counterparts among the catalogs are the same source, though discrepancies exist. For guidance, we include a ``cp\_coord\_flag'' to note which sources have counterparts with consistent coordinates and which do not, using a search radius of 2$^{\prime\prime}$ for SDSS, UKIDSS, VHS, and FIRST and 3$^{\prime\prime}$ for {\it WISE}, {\it GALEX}, and {\it Herschel} due to the larger PSF and higher astrometric uncertainties in these latter catalogs compared with the former. When the coordinates are inconsistent within these search radii, the ``cp\_coord\_flag'' is set to one, otherwise it is set to null. For 89\% of the X-ray sources with counterparts, their coordinates are consistent. We note, however, that above these search radii, consistent counterparts may exist and below these radii, there can still be discrepencies.

Finally, we highlight that the multi-wavelength magnitudes in the Stripe 82X catalogs may not be the most appropriate magnitude for every source and it is up to the user to determine whether different aperture photometry should be downloaded from the original catalog, using the identifying information presented in our catalogs to isolate the correct source, for the intended science goals.

A summary of the multi-wavelength columns and flags is presented in the Appendix, as well as a discussion of updates made to the previously released Stripe 82X catalogs.

\begin{deluxetable*}{lrrrrr}
\tablewidth{0pt}
\tablecaption{\label{multiwav_summary} Multi-Wavelength Counterpart Summary\tablenotemark{1}}
\tablehead{\colhead{Survey} & \colhead{{\it Chandra}} & \colhead{{\it
      XMM-Newton}} & \colhead{{\it XMM-Newton}} & \colhead{{\it
      XMM-Newton}} & \colhead{Total\tablenotemark{2}} \\
 & \colhead{Archival} & \colhead{Archival} & \colhead{AO10} &
 \colhead{AO13} } 

\startdata

SDSS\tablenotemark{3}  & 874 (118) &  1258 (190) & 614 (66) & 2438 (178) & 5009 (530) \\

{\it WISE}                        & 686 &  948 & 531& 2033 & 4006 \\

{\it UKIDSS}                    & 568 & 923 & 503 & 1784 & 3643 \\

VHS                                & 610 & 995 & 518 & 2119 & 4093 \\

{\it GALEX}                     & 166 & 254 &   82 &   607 & 1080 \\

FIRST                              &  42 &   55 &   27 &   116 &   232 \\

{\it Herschel}                  &  9   &     9 & \nodata & 121 & 133 \\

Redshifts                        & 339 & 465 & 292      & 828 & 1842
\enddata
\tablenotetext{1}{The counterpart numbers quoted in the text refer to associations found from matching the individual X-ray catalogs with the multi-wavelength source lists. Here, we include the final numbers that include ``promoted'' matches, found from cross-correlating the counterparts among the X-ray catalogs (see Sections \ref{ao13_xmatch} and \ref{arch_xmatch} for details).}
\tablenotetext{2}{Duplicate sources among surveys removed from
  total numbers.}
\tablenotetext{3}{Includes matches to the single-epoch and coadded
  catalogs. The number of sources found in the coadded catalog that do not have matches in the single-epoch data are
  quoted in parentheses.}
\end{deluxetable*}

\section{Discussion}

\begin{figure}
\centering
\includegraphics[angle=90,scale=0.35]{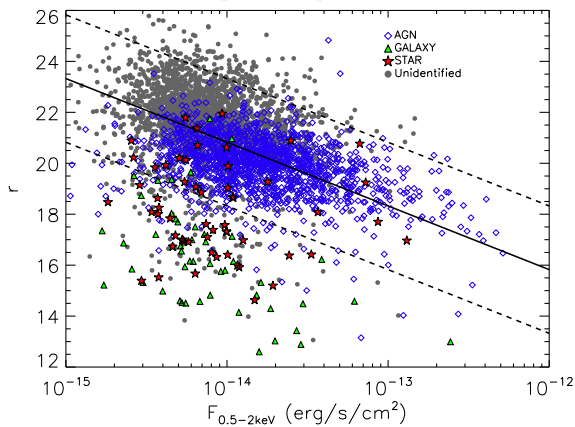}
\caption{\label{r_v_fx} SDSS $r$-band magnitude as a function of observed X-ray flux in the 0.5-2 keV band. The solid line defines the typical X-ray-to-optical flux ratio of AGN \citep{brandthas}, while the dashed lines show the X/O = $\pm$1 locus within which most AGN lie (see Equation \ref{xo}). Stars are identified by their optical spectra while AGN and galaxies are classified based on their observed 0.5-10 keV luminosity, with 10$^{42}$ erg s$^{-1}$ being the dividing line.}
\end{figure}

When considering the full Stripe 82X survey to date, including archival {\it Chandra}, archival {\it XMM-Newton}, {\it XMM-Newton} AO10, and {\it XMM-Newton} AO13 data, we find multi-wavelength counterparts to 88\% of the X-ray sources. We are able to identify $\sim$30\% of the Stripe 82X sample with spectroscopic objects. Sixty-seven objects are classified as stars while the remaining 1775 objects are extragalactic. We plot the $r$-band magnitude as a function of soft X-ray flux for the full sample in Figure \ref{r_v_fx}, where we note which objects are stars, X-ray AGN, X-ray galaxies, and currently unidentified (i.e., they lack redshifts). Stars are classified on the basis of their optical spectra while here we use the observed, full-band X-ray luminosity to differentiate between X-ray AGN ($L_{\rm 0.5-10keV} > 10^{42}$ erg s$^{-1}$) and X-ray galaxies ($L_{\rm 0.5-10keV} < 10^{42}$ erg s$^{-1}$), independent of their optical spectroscopic classification. For reference, we also include lines to mark typical AGN X/O values \citep[e.g.,][]{brandthas}:
\begin{equation}\label{xo}
{\rm X/O} = {\rm Log}(f_x/f_{\rm opt}) = log(f_x) + C + 0.4\times m_{\rm r},
\end{equation}
\noindent where $C$ is a constant based on the optical filter, which for the SDSS $r$-band, is 5.67 \citep[see][]{green_champ}. Previous studies have found that AGN generally fall within the X/0 = 0$\pm$1 locus \citep[e.g.,][]{schmidt,alexander,green_champ,brusa2007,xue,civano}, as indicated by the dashed lines in Figure \ref{r_v_fx}. We find the same trend here, and note that extragalactic objects do not separate out from Galactic objects within this color space.

\subsection{Stars}
\begin{figure}
\centering
\includegraphics[angle=90,scale=0.35]{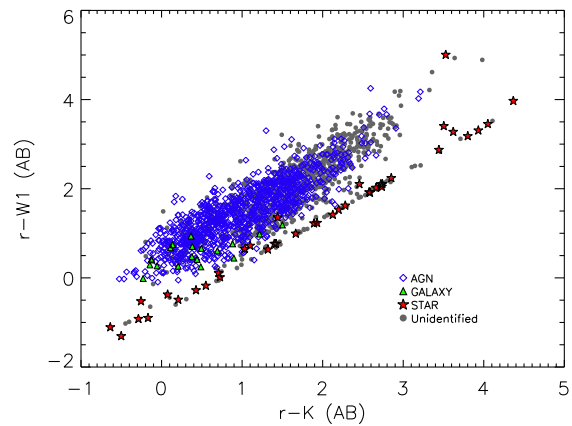}
\caption{\label{rw1_rk} $r-W1$ (AB) color as a function of $r-K$ (AB) color for the 1891 X-ray sources with SDSS, UKIDSS, and {\it WISE} counterparts that have $K$-band and $W1$ detections and UKIDSS ({\it WISE}) coordinates within 2$^{\prime\prime}$ (3$^{\prime\prime}$) of the SDSS position. Many of the stars can be identified by the distinct track they occupy in this color space \citep{r-w1}.}
\end{figure}

In Figure \ref{rw1_rk}, we show how most X-ray emitting stars can be cleanly identified on the basis of their optical and infrared properties by comparing their $r-K$ and $r-W1$ colors, as presented in \citet{r-w1}. Here, we focus on the X-ray sources with SDSS, UKIDSS, and WISE counterparts that have $K$-band detections, $W1$ detections ($W1$ SNR $>$2), and an $r$-band magnitude under 22.2 (the 95\% completeness limit for the single-epoch SDSS imaging catalog) to avoid artificially inflating the colors to redder values. Additionally, we only retain the sources where the SDSS and UKIDSS coordinates are consistent within 2$^{\prime\prime}$ and the SDSS and {\it WISE} coordinates agree within 3$^{\prime\prime}$ to minimize spurious associations. In total, 1891 objects are shown in Figure \ref{rw1_rk}, compared with the 4133 sources shown in the previous plot, which are sources detected in the $r$ and soft X-ray bands. Most of the stars follow a well-defined track in $r-K$ versus $r-W1$ color space, aiding in the separation of Galactic and extragalactic candidates detected in X-ray surveys in the absence of supporting spectroscopic information.

\subsection{Extragalactic Objects}
\begin{figure*}
\centering
\includegraphics[angle=90,scale=0.35]{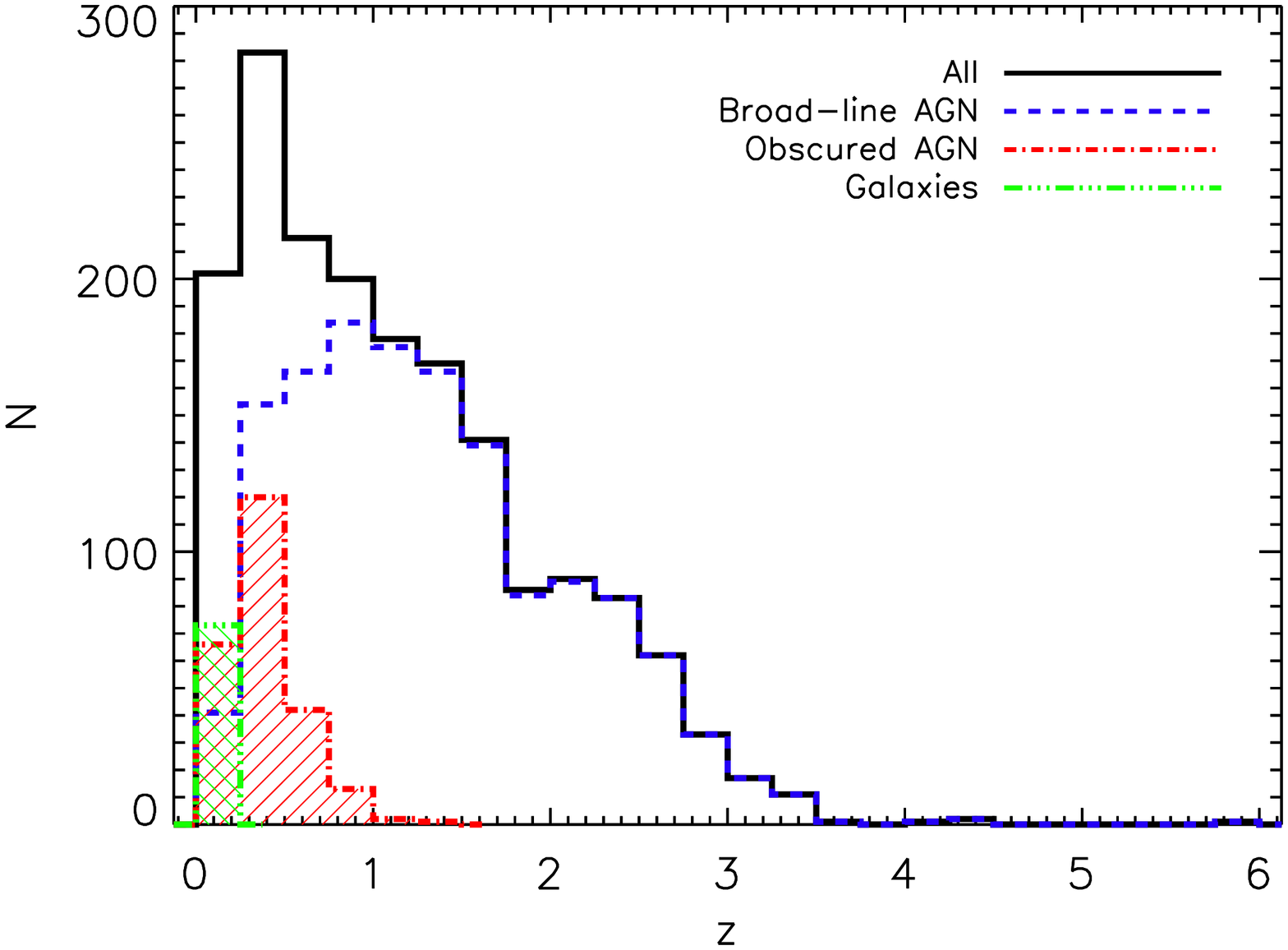}~
\includegraphics[angle=90,scale=0.35]{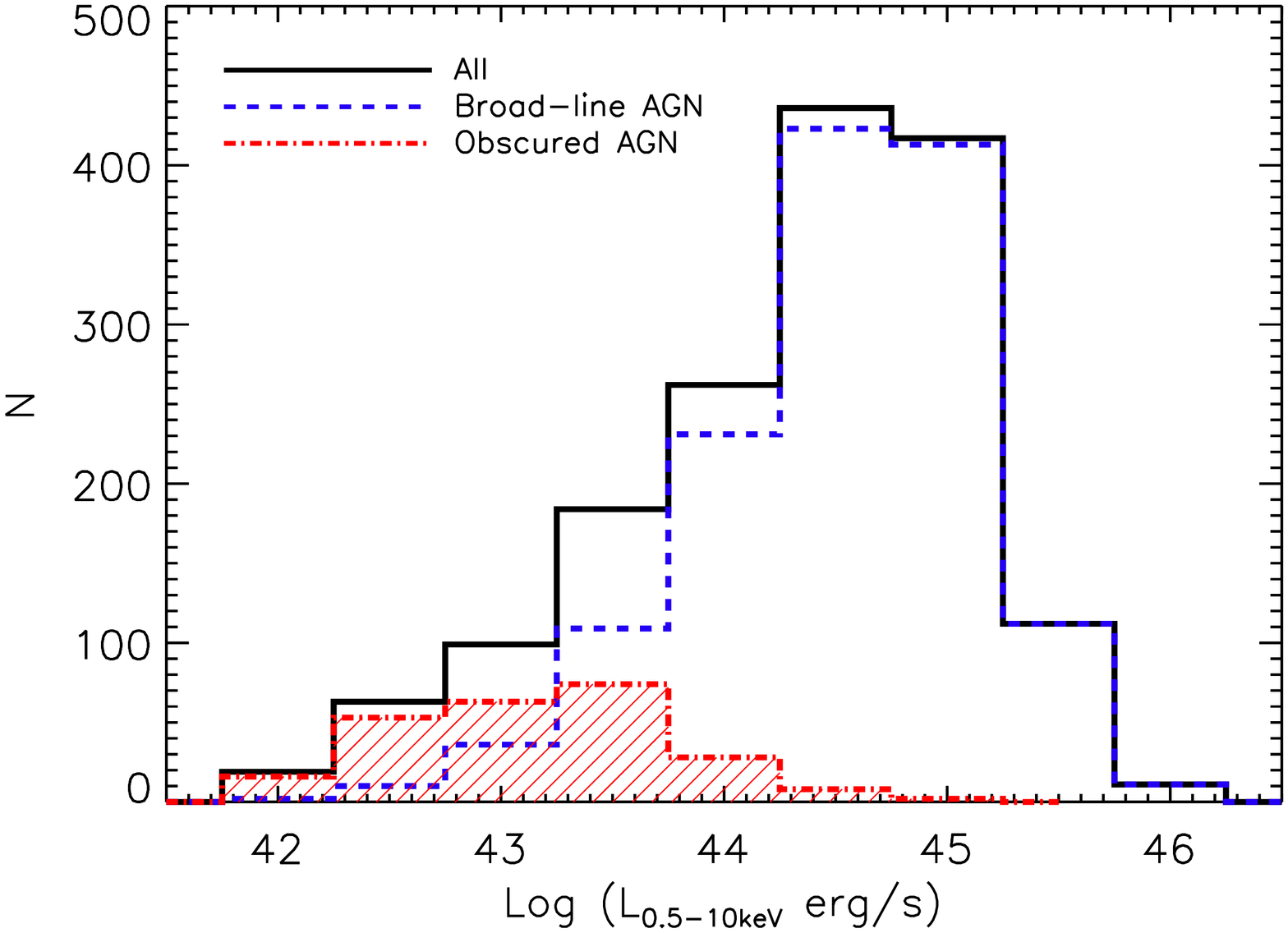}
\caption{\label{s82x_z_lum_hist} {\it Left}: Spectroscopic redshift distribution of the 1775 extragalactic Stripe 82X sources, with different classes of objects highlighted. Half the sample is above a redshift of one, and contains predominantly broad-line AGN at these distances. Nearly all obscured AGN (i.e., sources optically classified as galaxies with but with full band X-ray luminosities above 10$^{42}$ erg s$^{-1}$) are at a redshift below 1, while the optical and X-ray galaxies are at $z<0.25$ (Compton-thick AGN that have low observed X-ray flux due to heavy obscuration can be included in the ``galaxy'' bin). {\it Right}: Observed full-band luminosity distribution for the 1603 spectroscopically confirmed X-ray AGN (i.e., $L_{\rm 0.5-10keV} > 10^{42}$ erg s$^{-1}$), where the distribution peaks at high-luminosities (44.25 dex $<$ Log($L_{\rm 0.5-10keV}$ erg s$^{-1}$) $<$ 45.25 dex). High-luminosity AGN are predominantly broad-line sources while the lower-luminosity AGN are mostly obscured. We note that these trends are for the $\sim$30\% of the parent Stripe 82X sample that have spectroscopic redshifts and that with increased completeness and more sources identified via photometric redshifts, we expect to confirm more AGN at all luminosities and redshifts, including at $z>2$ and $L_{\rm 0.5-10keV} > 10^{45}$ erg s$^{-1}$, and a higher percentage of obscured AGN. }
\end{figure*}

In Figure \ref{s82x_z_lum_hist} (left), we show the redshift distribution of the 1775 extragalactic sources with optical spectra: about half (875) are at $z>1$, with 301 at redshifts above 2. We further break down the redshift distribution by classification, based on optical spectroscopy (see Section \ref{spectra}) and X-ray luminosity.  In Figure \ref{s82x_z_lum_hist}, ``broad-line'' AGN are sources optically classified as quasars due to broad emission lines in their spectra, ``obscured AGN'' are sources optically classified as galaxies whose full-band observed X-ray luminosities exceed 10$^{42}$ erg s$^{-1}$, and ``galaxies'' are objects lacking broad-lines in their optical spectra whose X-ray luminosities are below 10$^{42}$ erg s$^{-1}$; we note, however, that this ``galaxy'' class can include Compton-thick AGN ($N_H > 1.25\times10^{24}$ cm$^{-2}$) with very weak observed X-ray emission due to heavy attenuation. Of the 1775 extragalactic sources in our sample, 19 are not classified in the spectroscopic databases we utilized and another 30 do not have significant detections in the full X-ray band.

The left-hand panel of Figure \ref{s82x_z_lum_hist} demonstrates that nearly all the sources we have identified thus far at high redshifts (i.e., $z>1$) are broad-line AGN, in part because unobscured quasars were preferentially selected as spectroscopic targets in the SDSS surveys. Most of the obscured AGN live within the intermediate Universe ($z\sim0.5$) while galaxies reside in the local Universe ($z<0.25)$. We expect the percentage of obscured AGN, i.e., those lacking broad emission lines, to increase as more objects are identified via spectroscopic and photometric redshifts.

In the right-hand panel of Figure \ref{s82x_z_lum_hist}, we show the observed full-band luminosity distribution of the X-ray AGN, 1603 sources in total. The distribution peaks at relatively high luminosities ($\sim44.5$ dex) due to the wide-area and shallow design of the survey. Most of the high luminosity AGN are broad-line sources, though a handful of obscured AGN do reach moderately-high X-ray luminosities (Log ($L_{\rm 0.5-10keV}$/erg s$^{-1}$) $>43.75$ dex). 

\subsection{The $L-z$ Plane Probed by Stripe 82X}
\begin{figure*}
\centering
\includegraphics[angle=90,scale=0.35]{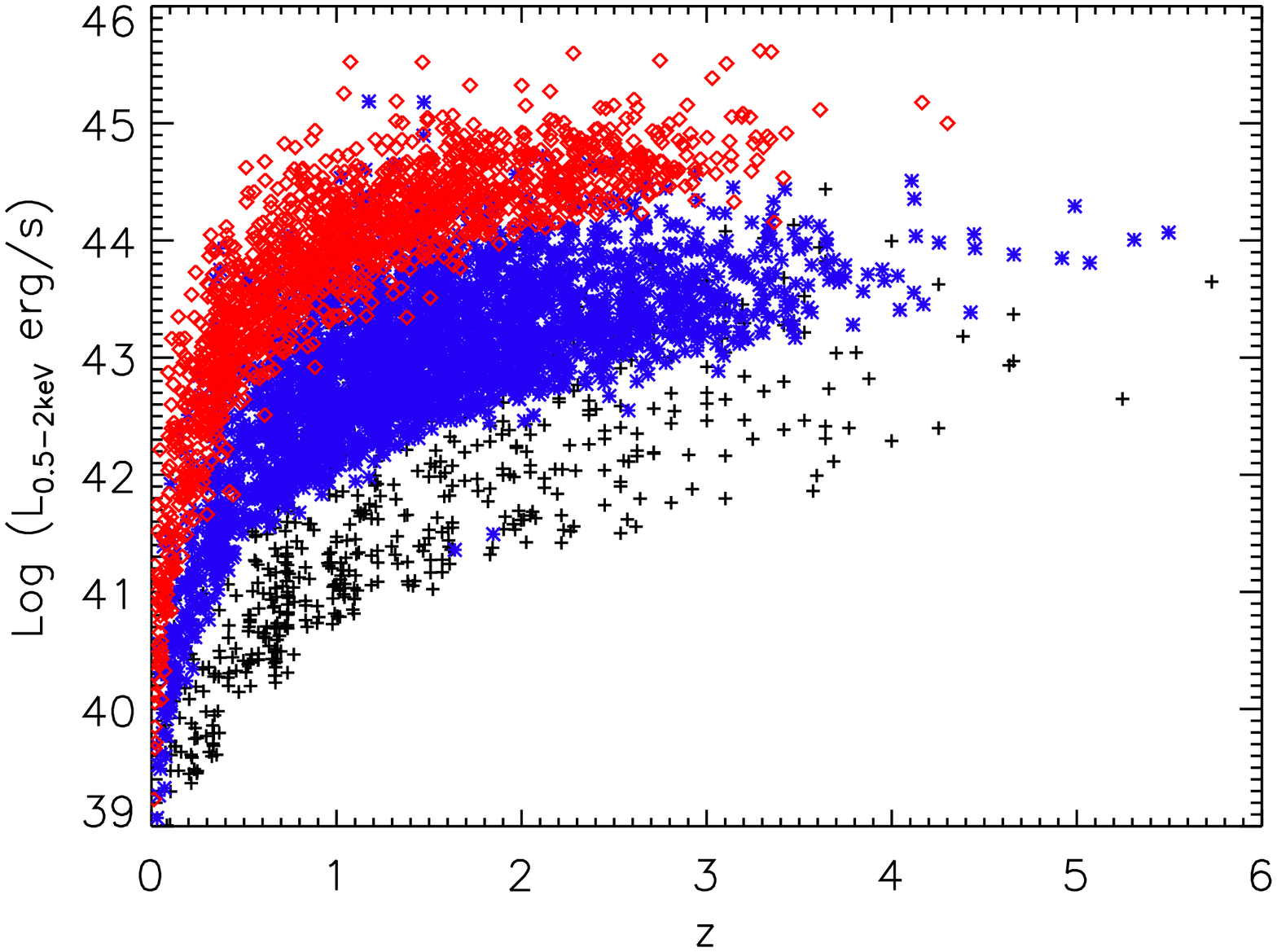}~
\includegraphics[angle=90,scale=0.35]{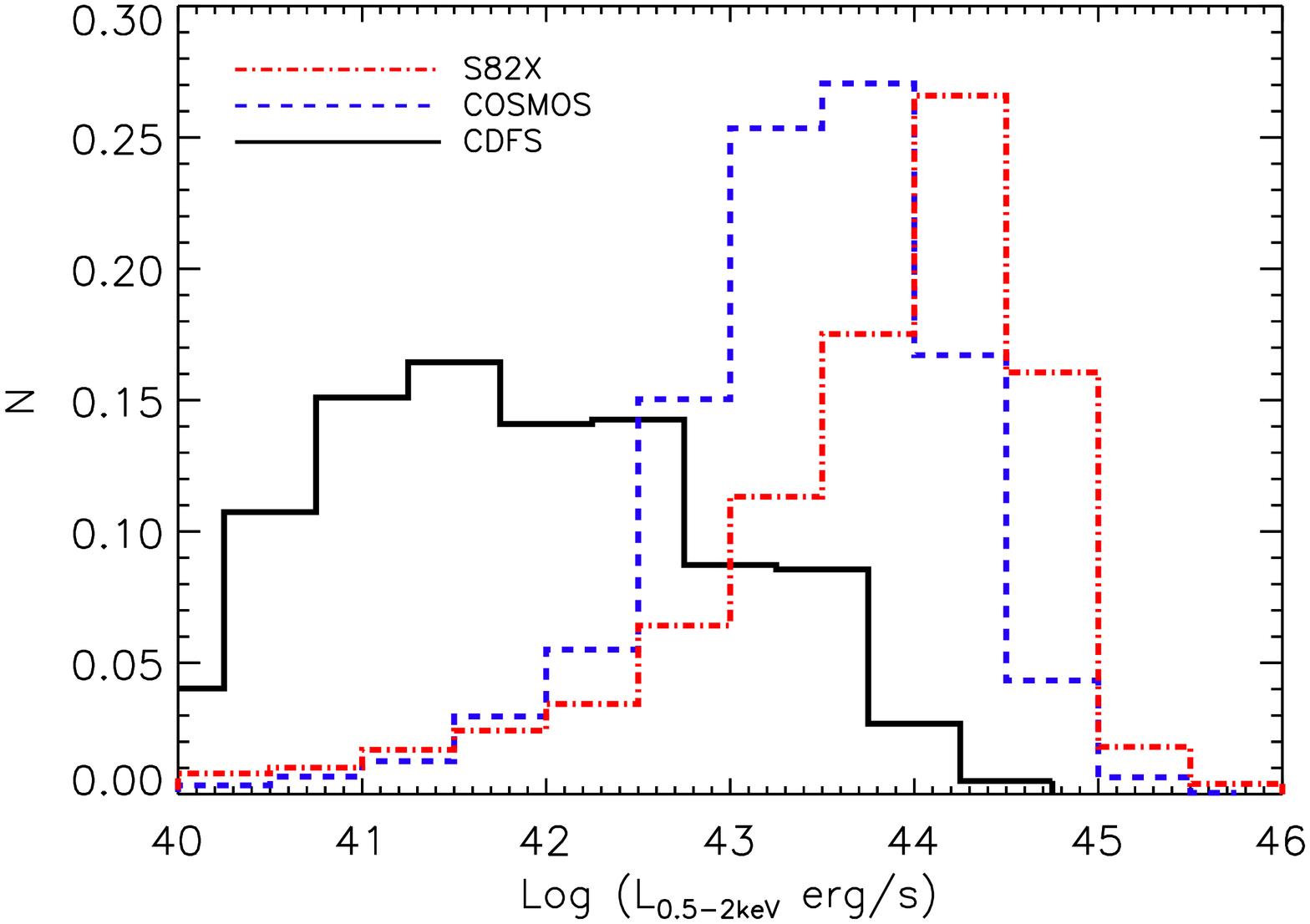}
\caption{\label{comp_lum} {\it Left}: $K$-corrected (rest-frame) soft-band (0.5-2 keV) luminosities as a function of redshift for the Stripe 82X (red diamonds), COSMOS-Legacy (blue asterisks; Civano et al. {\it submitted}, Marchesi et al. {\it submitted}), and CDFS (black crosses) sources. At every redshift, an increase in survey area preferentially identifies higher-luminosity sources. {\it Right}: Normalized distribution of $k$-corrected soft-band luminosities for Stripe 82X compared with COSMOS and CDFS: the wide-area coverage of Stripe 82X which probes a large effective volume of the Universe, enables the rare, highest luminosity quasars to be uncovered, complementing the parameter space explored by small- to moderate-area surveys. In both plots, only the sources identified with redshifts are plotted, representing 30\% of the Stripe 82X sample (which currently has only spectroscopic redshifts) and 91\% and 96\% of the CDFS and COSMOS-Legacy sample, respectively, where both spectroscopic and photometric redshifts are available.}
\end{figure*}

To put the Stripe 82X sample in context with other surveys, we compare the luminosity-redshift plane with the small-area, deep CDFS survey \citep[0.13 deg$^2$;][]{xue} and the moderate-area, moderate-depth COSMOS-Legacy survey (2.2 deg$^2$; Civano et al., {\it submitted}; Marchesi et al., {\it submitted}). Here, we use soft-band (0.5-2 keV) luminosities that have been $k$-corrected to the rest-frame, using $\Gamma$=1.4 for CDFS and COSMOS, while no $k$-correction was needed for Stripe 82X as the soft-band flux was estimated using $\Gamma$=2 and the $k$-correction scales as $(1+z)^{(\Gamma -2)}$.  As Figure \ref{comp_lum} (left) shows, as survey area increases and the effective flux limits of the surveys become shallower, the detected sources are preferentially at higher luminosity at every redshift. This is further illustrated in Figure \ref{comp_lum} (right), which compares the normalized luminosity distribution of Stripe 82X with COSMOS and CDFS, highlighting the complementarity of the different survey strategies in preferentially identifying sources within different luminosity ranges \citep[see, e.g.,][]{hsu}. Wide-area surveys which explore a large volume of the Universe, like Stripe 82X, are necessary to discover rare objects that have a low space density, including the highest luminosity quasars.

One important caveate in Figure \ref{comp_lum} is that we limit our comparison to sources with measured redshifts. For Stripe 82X, this represents the 30\% of the sample that has spectroscopic redshifts while COSMOS and CDFS have spectroscopic and photometric redshifts, effectively identifying $\sim$96\% and $\sim$91\% of the parent samples, respectively. The photometric redshift catalog for Stripe 82X will be published in the coming months (Ananna et al, {\it in prep.}), allowing us to identify the majority of the X-ray sources, increasing the number of sources at every redshift and luminosity.

\section{Conclusion}
We have presented the results from the most recent installment of the Stripe 82 X-ray survey, utilizing data awarded to our team in {\it XMM-Newton} cycle AO13.  This program, amounting to $\sim$980 ks of exposure time and spanning 15.6 deg$^2$, approximately doubled the previous X-ray coverage in the SDSS Stripe 82 Legacy field, with 2862 X-ray sources detected at $>$5$\sigma$ level. The approximate flux limits of the AO13 portion of the Stripe 82X survey are $2.2\times10^{-15}$ erg s$^{-1}$ cm$^{-2}$, $1.3\times10^{-14}$ erg s$^{-1}$ cm$^{-2}$, and $6.7\times10^{-15}$ erg s$^{-1}$ cm$^{-2}$, in the soft (0.5-2 keV), hard (2-10 keV), and full (0.5-10 keV) bands. From matching the X-ray source list to available multi-wavelength catalogs, including SDSS \citep{ahn,jiang}, {\it WISE} \citep{wright,mainzer}, UKIDSS \citep{hewett,lawrence,casali,warren}, VHS \citep{mcmahon}, {\it GALEX} \citep{morrissey}, FIRST \citep{becker2012,helfand}, and {\it Herschel} \citep{viero}, we identified reliable counterparts for 93\% of the sample. About 29\% of the X-ray sources are classified via spectroscopic redshifts.

Merging this dataset with our previous releases of the Stripe 82X catalogs \citep{me1,me2}, the X-ray survey area in Stripe 82 covers $\sim$31.3 deg$^2$, with 6181 unique X-ray sources detected at $\geq 4.5\sigma$ and $> 5\sigma$, for the {\it Chandra} and {\it XMM-Newton} components of the survey, respectively. We also updated the multi-wavelength matched X-ray catalogs for these earlier segments of the survey. In total, we find reliable multi-wavelength counterparts for 88\% of the full Stripe 82X survey to date, with a spectroscopic completeness of 30\%. We emphasize that we matched the X-ray source list to each multi-wavelength catalog independently and have not cross-correlated the counterparts. However the counterparts largely agree among the catalogs, as discussed in the main text. Care must also be taken when studying the colors or spectral energy distributions of the X-ray sources using the magnitudes we present in these catalogs: it is up to the user to decide whether the aperture photometry in these catalogs is most suitable for a given source or if different aperture magnitudes should be retrieved from the main multi-wavelength catalogs, using the identifying information in the Stripe 82X catalog to select specific sources.

The large volume of the Universe explored by the Stripe 82X survey enables the discovery of high-luminosity, high-redshift AGN, a missing tier in the current X-ray census of supermassive black hole growth. We have several upcoming dedicated spectroscopic follow-up programs to increase the completeness of Stripe 82X, which in tandem with the photometric redshift catalog (Ananna et al., {\it in prep.}), will allow us to constrain how the most luminous X-ray AGN evolve over cosmic time. Furthermore, we are targeting obscured AGN candidates which have optical and infrared clues that they may be the transitional link in the merger-induced black hole growth/galaxy evolution paradigm \citep[see, e.g.,][]{glikman2013,brusa2015}; a handful of such luminous obscured AGN at $z>1$ have already been discovered (LaMassa et al., {\it in prep.}), with many more candidates yet to be explored in this dataset. Additionally, Stripe 82X will provide insight into AGN host galaxies via spectral energy distribution analysis, AGN variability, the dark matter halos hosting high-luminosity quasars at high-redshift via clustering analysis, the X-ray properties of galaxy clusters, and the opportunity to discover exotic sources. For instance, in the previous release of the Stripe 82X catalog, we discovered the first ``changing-look'' quasar \citep{me_changinglook}, an AGN which transitioned from a broad-line (Type 1) object to a mostly narrow-line (Type 1.9) object within a 10 year period \citep[see also][]{merloni}. We expect that Stripe 82X will have applications beyond those listed here, and can be particularly helpful in informing best-effort practices for AGN identification and follow-up in upcoming wide-area surveys, including {\it eROSITA} which will be launched in 2017 \citep{merloni2012,erosita}. Finally, we note that increasing the X-ray area to 100 deg$^2$ will open a new window into black hole growth at $z>3$ and luminosities greater than 10$^{45}$ erg s$^{-1}$, which is only beginning to be explored from an X-ray perspective \citep[e.g.,][]{georgakakis}. The existing ancillary data will allow these objects to be readily characterized, allowing this population to be fully understood. 

\acknowledgements
We thank the anonymous referee for feedback that helped improve the manuscript. SML acknowledges support from grant number NNX15AJ40G. CMU gratefully acknowledges support from Yale University. CMU and SK would like to thank the Kavli Institute for Theoretical Physics (Santa Barbara) for their hospitality and support; the KITP is supported by  NSF Grant No. NSF PHY11-25915. MB acknowledges support from the FP7 Career Integration Grant “eEASy” (CIG 321913). Support for the work of ET was provided by the Center of Excellence in Astrophysics and Associated Technologies (PFB 06), by the FONDECYT grant 1120061 and the Anillo project ACT1101. KS gratefully acknowledges support from Swiss National Science Foundation Grant PP00P2\_138979/1.

Funding for the SDSS and SDSS-II has been provided by the Alfred P. Sloan Foundation, the Participating Institutions, the National Science Foundation, the U.S. Department of Energy, the National Aeronautics and Space Administration, the Japanese Monbukagakusho, the Max Planck Society, and the Higher Education Funding Council for England. The SDSS Web Site is http://www.sdss.org/.

The SDSS is managed by the Astrophysical Research Consortium for the Participating Institutions. The Participating Institutions are the American Museum of Natural History, Astrophysical Institute Potsdam, University of Basel, University of Cambridge, Case Western Reserve University, University of Chicago, Drexel University, Fermilab, the Institute for Advanced Study, the Japan Participation Group, Johns Hopkins University, the Joint Institute for Nuclear Astrophysics, the Kavli Institute for Particle Astrophysics and Cosmology, the Korean Scientist Group, the Chinese Academy of Sciences (LAMOST), Los Alamos National Laboratory, the Max-Planck-Institute for Astronomy (MPIA), the Max-Planck-Institute for Astrophysics (MPA), New Mexico State University, Ohio State University, University of Pittsburgh, University of Portsmouth, Princeton University, the United States Naval Observatory, and the University of Washington.

Funding for SDSS-III has been provided by the Alfred P. Sloan Foundation, the Participating Institutions, the National Science Foundation, and the U.S. Department of Energy Office of Science. The SDSS-III web site is http://www.sdss3.org/.

SDSS-III is managed by the Astrophysical Research Consortium for the Participating Institutions of the SDSS-III Collaboration including the University of Arizona, the Brazilian Participation Group, Brookhaven National Laboratory, Carnegie Mellon University, University of Florida, the French Participation Group, the German Participation Group, Harvard University, the Instituto de Astrofisica de Canarias, the Michigan State/Notre Dame/JINA Participation Group, Johns Hopkins University, Lawrence Berkeley National Laboratory, Max Planck Institute for Astrophysics, Max Planck Institute for Extraterrestrial Physics, New Mexico State University, New York University, Ohio State University, Pennsylvania State University, University of Portsmouth, Princeton University, the Spanish Participation Group, University of Tokyo, University of Utah, Vanderbilt University, University of Virginia, University of Washington, and Yale University.

This publication makes use of data products from the {\it Wide-field Infrared Survey Explorer}, which is a joint project of the University of California, Los Angeles, and the Jet Propulsion Laboratory/California Institute of Technology, funded by the National Aeronautics and Space Administration.

\appendix
\section{Updates to Previously Released Stripe 82X Catalogs}\label{update}
As mentioned in Section \ref{multi_wav}, the MLE matching between the archival {\it XMM-Newton} and AO10 source lists and ancillary catalogs was updated to include a 1$^{\prime\prime}$ systematic error added in quadrature to the {\it emldetect} reported positional error. Since the {\it Chandra} Source Catalog has external astrometric corrections applied \citep{rots}, a systematic positional error did not need to be added in by hand for this source list. Additional updates to the previously published catalogs are listed below, where Table \ref{multiwav_summary} summarizes the multi-wavelength catalog matching for all components of the Stripe 82X survey.

\subsection{X-ray Catalogs}
In the X-ray source lists for both the {\it XMM-Newton} and {\it Chandra} catalogs, we now include columns for net counts detected in the soft, hard, and full bands. We also updated the {\it XMM-Newton} catalog to include an ``ext\_flag'' whereas the previous version had the fluxes in the band that were fit as extended by {\it emldetect} set to zero; these fluxes now reflect the values reported by {\it emldetect}. Additionally, in the previous {\it XMM-Newton} catalog, we set to null any flux that was not detected at the {\it det\_ml}$\geq 15$ level. Here, we report the fluxes along with their corresponding {\it det\_ml} value for the user to decide which flux significance is most appropriate for their purposes. For fluxes from the {\it Chandra} source list, we determined the 4.5$\sigma$ significance by comparing the catalog with the sensitivity maps \citep[see][]{me1}, nulling out any fluxes which were below this significance threshold; we refer the user to the {\it Chandra} Source Catalog \citep{evans} for flux measurements at lower significance. 

The ``removed\_logn\_logs'' flags have been updated in both catalogs to indicate which sources were excluded from the Log$N$-Log$S$ generation in this work. These discarded sources represent those that are in overlapping observations that were excluded from the area-flux curve or that were targeted sources in archival observations.

Finally, in the previous catalog releases, we noted whether a {\it Chandra} source was found in the {\it XMM-Newton} catalog and vice versa, using the matching algorithm discussed above to find matches between the {\it XMM-Newton} source lists generated via {\it emldetect}. We added a column to note whether a source is also detected in the {\it XMM-Newton} AO13 catalog, as well as the unique identifying information for the matched source (i.e., the {\it Chandra} MSID if the ``in\_chandra'' flag is set to ``yes'' and the {\it XMM-Newton} record number if the ``in\_xmm'' or ``in\_xmm\_ao13'' flag is set to ``yes'').

\subsection{Coadded SDSS Catalog} 
We followed the same procedure detailed above when matching the previously released X-ray catalogs to the coadded SDSS source list. Again, $r_{\rm search}$ is 5$^{\prime\prime}$ and 7$^{\prime\prime}$ for {\it Chandra} and {\it XMM-Newton} respectively \citep{civano,brusa2010,me2}. The $R_{\rm crit}$ values when matching to the {\it Chandra} source list are as follows: $u$ - 0.7, $g$ - 0.9, $r$ - 0.85, $i$ - 0.85, and $z$ 0.8, with the fraction of random to true matches above this threshold being 14/572, 16/543, 19/601, 18/601, and 13/816, respectively. For {\it XMM-Newton}, we impose $R_{\rm crit}$ values of 0.85, 0.9, 0.9, 0.9, 0.85 in the $u$, $g$, $r$, $i$, and $z$ bands, respectively, with spurious fractions of 29/1290, 39/1326, 60/1283, 54/1228, 49/1420. We retain the coadded match if there is not a counterpart found in the single-epoch imaging. By matching to the coadded catalog, we gain additional optical counterparts to 139 of the {\it Chandra} sources and 250 of the {\it XMM-Newton} sources. 

\subsection{Optical Spectroscopy}
Since publishing the previous release of the Stripe 82X catalog, we have mined these additional databases for spectroscopic redshifts: PRIMUS \citep{coil}, the \citet{ross_mmt} pre-BOSS pilot survey with Hectospec, 6dF \citep{jones2004,jones2009},  and VVDS \citep{garilli}. We also have an on-going ground-based follow-up campaign to target X-ray sources and have redshifts from WIYN HYDRA from observing runs in 2012 December, 2013 August-September, 2014 January, 2014 June, 2014 July, 2014 September, and 2015 January; from ISAAC on VLT from 2013 August; from NIRSPEC on Keck 2013 September; and from Palomar DoubleSpec from 2014 July and 2014 December.  We now have redshifts, and where available, optical classifications for 335 of the {\it Chandra} sources and 760 of the {\it XMM-Newton} sources; 142 of these redshifts are from our follow-up observing program, where the WIYN HYDRA spectra were reduced with the {\it IRAF} routine {\it dohydra}, the ISAAC spectrum was reduced with the VLT provided {\it esorex} pipeline, the NIRSPEC data were reduced with the IRAF task {\it wmkonspec}, and the Palomar DoubleSpec spectra were extracted using the conventional single-slit extraction routines in {\it IRAF}. Stars and extragalactic objects were classified on the basis of their emission and/or absorption features.

\subsection{AllWISE Catalog}
We now match the archival {\it Chandra}, archival {\it XMM-Newton}, and AO10 {\it XMM-Newton} source lists to the AllWISE rather than AllSky Catalog, superseding the matches published in \citet{me2}. We impose a $R_{\rm crit}$ cut of 0.7 for the {\it Chandra}/AllWISE matching, finding 5 spurious associations out of 700. In total, 667 {\it WISE} sources survived the quality control cuts, where all 33 rejected sources were extended with either null or flagged extended photometry; two extended sources had acceptable elliptical aperture photometry measurements in at least one band. For the archival and AO10 {\it XMM-Newton} matching, our $R_{\rm crit}$ value is 0.85, with a spurious fraction of 20/1516.  We were left with 1465 {\it WISE} sources that passed the quality control checks, of which four were extended. Forty-eight extended sources were rejected as were three point sources. For reference, matching to the AllSky {\it WISE} catalog garnered 595 and 1398 {\it Chandra} and {\it XMM-Newton} sources, respectively, so we increase the percentage of X-ray sources with {\it WISE} counterparts from 52\% to 58\% and 59\% to 62\%, respectively. We note that the archival X-ray data have deeper pointings, causing the association rate to be lower than for the {\it XMM-Newton} AO13 source list which is at a relatively shallow depth.

\subsection{VHS}
Similar to the {\it XMM-Newton} AO13 catalog, we include columns for matching the archival {\it Chandra},  archival {\it XMM-Newton}, and AO10 {\it XMM-Newton} source lists to the VHS catalog. We find critical $R_{\rm crit}$ values of 0.85 for $J$ and 0.8 for $H$ and $K$ when matching to the {\it Chandra} source list, with an estimated 5/530, 6/500, and 8/544 contamination rate in the $J$, $H$, and $K$ bands, respectively. In total, 577 VHS counterparts are found for the {\it Chandra} sources (50\% of the sample), with none rejected for being saturated. Between UKIDSS and VHS, there are NIR counterparts for 689 {\it Chandra} sources, or 60\% of the source list. Of the 112 X-ray/UKIDSS objects without a VHS counterpart, 41 are undetected in the VHS survey; the remaining were below the $R_{\rm crit}$ threshold.

When matching the archival and AO10 {\it XMM-Newton} catalog the VHS source list, we used a $R_{\rm crit}$ threshold of 0.75 in the $J$ and $H$ bands and 0.8 in the $K$ band, with  spurious association rates of 30/1250, 27/1200, and 38/1280, respectively. We find 1504 VHS counterparts to the {\it XMM-Newton} sources (64\% of the sample), while 1670 X-ray objects (71\%) have NIR matches in either UKIDSS or VHS. Of the 166 X-ray/UKIDSS sources without a VHS match, 45 were undetected in the VHS survey.

\subsection{Herschel}
Several of the archival {\it Chandra} and {\it XMM-Newton} observations overlap the HerS survey area \citep{viero}, so we use a nearest neighbor match to find counterparts to these X-ray sources. Again, the {\it XMM-Newton} search radius is 5$^{\prime\prime}$. In \citet{me2}, we used a 5$^{\prime\prime}$ search radius to find counterparts to {\it Chandra} sources, though here we lower this search radius to 3$^{\prime\prime}$ when matching to {\it Herschel} since our exercise of matching the randomized X-ray source positions to the {\it Herschel} catalog found false matches only at radii above 5$^{\prime\prime}$. For both the {\it Chandra} and {\it XMM-Newton} matches to {\it Herschel}, no associations were found between the randomized X-ray positions and the {\it Herschel} source list.

\subsection{Catalog Cross-Matches}
\label{arch_xmatch}
The previously released versions of the Stripe 82X multi-wavelength matched catalogs did not include promoted matches found from cross-correlating the individual catalogs as discussed above. In the current versions of the catalogs, the promoted matches are included as well as flags to indicate which multi-wavelength counterparts are added into the catalog in this manner.

Additionally, we also have included a ``cp\_coord\_flag,'' as described in the main text. The coordinates among the multi-wavelength counterparts are consistent for 96\% of the {\it Chandra} sources and for 92\% of the {\it XMM-Newton} sources.

\section{Catalog Column Summary}

\FloatBarrier
\begin{table}
\caption{Common Columns Among all X-ray Catalogs: X-ray Information}
\begin{tabular}{p{0.3\linewidth}p{0.6\linewidth}}
\hline
\noalign{\smallskip}\hline\noalign{\smallskip}
Column & Description \\
\noalign{\smallskip}\hline \noalign{\smallskip}

ObsID & {\it Chandra} or {\it XMM-Newton} observation identification number. \\

RA       & X-ray RA (J2000). \\

Dec     &  X-ray Dec (J2000). \\

RADec\_err & Positional error on the X-ray coordinates in arcseconds. \\

Dist\_NN & Distance to the nearest X-ray source in the catalog in arcseconds. \\

Removed\_LogN\_LogS ({\it Chandra and XMM-Newton archival and AO10 catalogs only}) & Flag set to ``yes'' if removed from the LogN-LogS relations presented here. The removed objects are targeted sources of observations and, in the case of the {\it Chandra} catalog, objects that overlap {\it XMM-Newton} observations in the field. \\

Soft\_Flux & Observed X-ray flux in the soft (0.5-2 keV) band (erg s$^{-1}$ cm$^{-2}$). For the {\it Chandra} sources, only significant ($\geq 4.5 \sigma$) fluxes are reported (see text for details) while all fluxes are reported for the {\it XMM-Newton} sources, where users should refer to the ``soft\_detml'' entry to determine appropriate level of source significance suitable for analysis. Fluxes are converted from count rate assuming a powerlaw spectrum where $\Gamma$=2.0. \\

Soft\_Counts & Net counts in the soft (0.5-2 keV) band. \\

Hard\_Flux & Observed X-ray flux in the hard band, which corresponds to the 2-7 keV range for {\it Chandra} and 2-10 keV range for {\it XMM-Newton} (erg s$^{-1}$ cm$^{-2}$). For the {\it Chandra} sources, only significant ($\geq 4.5 \sigma$) fluxes are reported (see text for details) while all fluxes are reported for the {\it XMM-Newton} sources, where users should refer to the ``hard\_detml'' entry to determine appropriate level of source significance suitable for analysis. Fluxes are converted from count rate assuming a powerlaw spectrum where $\Gamma$=1.7.\\

Hard\_Counts & Net counts in the hard band (2-7 keV and 2-10 keV for {\it Chandra} and {\it XMM-Newton} respectively). \\

Full\_Flux &  Observed X-ray flux in the full band, which corresponds to the 0.5 - 7 keV range for {\it Chandra} and 0.5 - 10 keV range for {\it XMM-Newton} (erg s$^{-1}$ cm$^{-2}$). For the {\it Chandra} sources, only significant ($\geq 4.5 \sigma$) fluxes are reported (see text for details) while all fluxes are reported for the {\it XMM-Newton} sources, where users should refer to the ``full\_detml'' entry to determine appropriate level of source significance suitable for analysis. Fluxes are converted from count rate assuming a powerlaw spectrum where $\Gamma$=1.7.\\

Full\_Counts &  Net counts in the full band (0.5-7 keV and 0.5-10 keV for {\it Chandra} and {\it XMM-Newton} respectively). \\

Lum\_Soft & Log of the observed luminosity in the soft (0.5-2 keV) band in units of erg s$^{-1}$. \\

Lum\_Hard & Log of the observed luminosity in the hard band (2-7 keV and 2-10 keV for {\it Chandra} and {\it XMM-Newton} respectively), in units of erg s$^{-1}$. \\

Lum\_Full &  Log of the observed luminosity in the full band (0.5-7 keV and 0.5-10 keV for {\it Chandra} and {\it XMM-Newton} respectively), in units of erg s$^{-1}$. \\
\noalign{\smallskip}\hline \noalign{\smallskip}
\end{tabular}
\end{table}

\clearpage

\begin{longtable} {p{0.3\linewidth}p{0.6\linewidth}}
\caption[]{Common Columns Among all Catalogs: Multi-wavelength Information}\\
\hline 
\noalign{\smallskip}\hline\noalign{\smallskip}
Column & Description \\
\noalign{\smallskip}\hline\noalign{\smallskip}
\endfirsthead

\multicolumn{2}{l}{{\bfseries \tablename \thetable{} -- continued from previous page}}\\
\hline
\noalign{\smallskip}\hline\noalign{\smallskip}
Column & Description \\
\noalign{\smallskip}\hline\noalign{\smallskip}
\endhead

\noalign{\smallskip}\hline\noalign{\smallskip} \multicolumn{2}{r}{{Continued on next page}} \\ \noalign{\smallskip}\hline\noalign{\smallskip}
\endfoot

\hline 
\endlastfoot

SDSS\_rej & Flag set to ``yes'' if SDSS counterpart found which exceeds reliability threshold, but the photometry was rejected for failing quality control checks. \\

SDSS\_OBJID & SDSS object identification number of SDSS counterpart to X-ray source. \\

SDSS\_RA & SDSS RA of counterpart (J2000). \\

SDSS\_Dec & SDSS Dec of counterpart (J2000). \\

SDSS\_Rel &  MLE reliability value of SDSS counterpart; highest of the $u$, $g$, $r$, $i$, and $z$ reliability values. \\

SDSS\_Dist &  Distance between X-ray source and SDSS counterpart in arcseconds. \\

SDSS\_Coadd & Flag set to ``yes'' if SDSS counterpart found from the \citet{jiang} coadded catalog. Otherwise, the SDSS counterpart was identified in the single-epoch DR9 imaging catalog. \\

u\_mag &  SDSS $u$-band magnitude. For the single-epoch matches, this value represents the SDSS pipeline reported ModelMag while the sources from the coadded SDSS catalog have Mag\_Auto values calculated via SExtractor \citep{bertin} as reported in the \citet{jiang} catalogs. \\

u\_err &  SDSS $u$-band magnitude error. For the single-epoch matches, this value represents the SDSS pipeline reported ModelMagErr while the sources from the coadded SDSS catalog have MagErr\_Auto values calculated via SExtractor \citep{bertin} as reported in the \citet{jiang} catalogs. \\ 

g\_mag & SDSS $g$-band magnitude.  See $u$\_mag for details. \\

g\_err  & SDSS $g$-band magnitude error. See $u$\_err for details.  \\ 

r\_mag & SDSS $r$-band magnitude. See $u$\_mag for details. \\ 

r\_err & SDSS $r$-band magnitude error. See $u$\_err for details. \\ 

i\_mag & SDSS $i$-band magnitude. See $u$\_mag for details. \\

i\_err & SDSS $i$-band magnitude error. See $u$\_err for details. \\ 

z\_mag &  SDSS $z$-band magnitude. See $u$\_mag for details. \\

z\_err & SDSS $z$-band magnitude error. See $u$\_err for details. \\ 

Specobjid &  SDSS spectroscopic identification number.\\

Class & Optical spectroscopic class as indicated by pipeline processing (for spectra downloaded from existing databases) or determined by us through visual inspection for sources targeted in our follow-up campaigns. QSO - broad emission lines in spectra; GALAXY - narrow emission lines and/or absorption lines only; STAR - stellar spectrum. \\

Redshift & Spectroscopic redshift. \\

z\_src & Source of spectroscopic redshift: 0 - SDSS DR9 \citep{ahn}; 1 - 2SLAQ \citep{croom}; 2 - WiggleZ \citep{drinkwater}; 3 - DEEP2 \citep{newman}; 4 - sources with ``ZWARNING'' flag set in SDSS pipeline which were visually inspected by us where the redshift was confirmed, updated, or nulled out if no solution could be found; 5 - SDSS DR10 \citep{dr10}; 6 - the spectroscopic survey of faint quasars in Stripe 82 from \citet{jiang2006}; 7 - WIYN HYDRA follow-up observations on 2012 Dec; 8 - PRIMUS \citep{coil}; 9 - VLT ISAAC follow-up observation on 2013 August; 10 - Keck NIRSPEC follow-up observations on 2013 September; 11 - SDSS DR12 \citep{dr12}; 12 - WIYN HYDRA follow-up observations on 2013 August - Sep; 13 - WIYN HYDRA follow-up observations on 2014 January; 14 - WIYN HYDRA follow-up observations on 2014 June; 15 - WIYN HYDRA follow-up observations on 2014 July; WIYN HYDRA follow-up observations on 2014 September; 17 - WIYN HYDRA follow-up observations on 2015 January; 18 - Palomar DoubleSpec observations on 2014 July;  19 - pre-BOSS pilot survey using Hectospec on MMT \citep{ross_mmt}; 20 - Palomar DoubleSpec follow-up observations on 2014 December; 22 - 6dF \citep{jones2004,jones2009}; 23 - VVDS \citep{lefevre2003,lefevre2005,garilli,lefevre2013}. \\

WISE\_Name &  {\it WISE} name.\\

WISE\_RA &  RA of {\it WISE} counterpart (J2000). \\

WISE\_Dec & Dec of {\it WISE} counterpart (J2000). \\

WISE\_sigra &  Uncertainty of {\it WISE} RA (arcsec). \\

WISE\_sigdec & Uncertainty of {\it WISE} Dec (arcsec). \\

WISE\_Rel &  MLE reliability value of {\it WISE} counterpart. \\

WISE\_Dist &  Distance between {\it WISE} counterpart and X-ray source (arcsec). \\

W1 &$W1$ magnitude (Vega). \\

W1sig & $W1$ error. \\

W1SNR & $W1$ signal-to-noise ratio. Magnitudes with SNR $<$ 2 are upper limits.\\

W2 & $W2$ magnitude (Vega). \\

W2sig &  $W2$ error. \\

W2SNR & $W2$ signal-to-noise ratio. Magnitudes with SNR $<$ 2 are upper limits.\\

W3 &  $W3$ magnitude (Vega). \\

W3sig &  $W3$ error. \\

W3SNR &   $W3$ signal-to-noise ratio. Magnitudes with SNR $<$ 2 are upper limits.\\

W4 &  $W4$ magnitude (Vega). \\

W4sig & $W4$ error. \\

W4SNR &  $W4$ signal-to-noise ratio. Magnitudes with SNR $<$ 2 are upper limits.\\

WISE\_ext &  Flag set to ``yes'' if the {\it WISE} source is extended. \\

WISE\_rej &  Flag set to ``yes'' if {\it WISE} counterpart is identified via MLE matching but the source is rejected due to failing photometry control checks in every {\it WISE} band.\\

UKIDSS\_ID &  Identification number of UKIDSS counterpart.\\

UKIDSS\_RA &  RA of UKIDSS counterpart (J2000). \\

UKIDSS\_Dec &  Dec of UKIDSS counterpart (J2000). \\

UKIDSS\_Rel &  MLE reliability value of UKIDSS counterpart; highest of the $Y$, $J$, $H$, and $K$ reliability values. \\

UKIDSS\_Dist &  Distance between UKIDSS counterpart and X-ray source (arcsec). \\

UKIDSS\_Ymag &  UKIDSS $Y$ magnitude (Vega). \\

UKIDSS\_Ysig & UKIDSS $Y$ magnitude error (Vega). \\

UKIDSS\_Jmag & UKIDSS $J$ magnitude (Vega). \\

UKIDSS\_Jsig &  UKIDSS $J$ magnitude error (Vega). \\

UKIDSS\_Hmag &  UKIDSS $H$ magnitude (Vega). \\

UKIDSS\_Hsig &  UKIDSS $H$ magnitude error (Vega). \\

UKIDSS\_Kmag &  UKIDSS $K$ magnitude (Vega). \\

UKIDSS\_Ksig &  UKIDSS $K$ magnitude error (Vega). \\

UKIDSS\_rej & flag set to ``yes'' if UKIDSS counterpart is found via MLE matching but source is rejected due to failing quality control checks.\\

VHS\_ID & Identification number of VHS counterpart.\\

VHS\_RA & RA of VHS counterpart (J2000). \\

VHS\_Dec & Dec of VHS counterpart (J2000). \\

VHS\_Rel &  MLE reliability value of VHS counterpart; highest of the $J$, $H$, and $K$ reliability values. \\

VHS\_Dist & Distance between VHS counterpart and X-ray source (arcsec). \\

VHS\_Jmag & VHS $J$ magnitude (Vega). \\

VHS\_Jsig & VHS $J$ magnitude error (Vega). \\

VHS\_Hmag & VHS $H$ magnitude (Vega). \\

VHS\_Hsig & VHS $H$ magnitude error (Vega). \\

VHS\_Kmag & VHS $K$ magnitude (Vega). \\

VHS\_Ksig & VHS $K$ magnitude error (Vega). \\

VHS\_rej & flag set to ``yes'' if VHS counterpart is found via MLE matching but source is rejected due to failing quality control checks.\\

GALEX\_RA & RA of {\it GALEX} counterpart (J2000). \\

GALEC\_Dec & Dec of {\it GALEX} counterpart (J2000). \\

NUV\_poserr &  Positional error on the NUV {\it GALEX} position (arcsec). \\

FUV\_poserr &  Positional error on the FUV {\it GALEX} position (arcsec). \\

GALEX\_Rel &  MLE reliability value of the {\it GALEX} counterpart; higher of the FUV and NUV reliability values. \\

GALEX\_Dist &  Distance between {\it GALEX} counterpart and X-ray source (arcsec). \\

NUV\_Mag &  NUV magnitude (AB). \\

NUV\_MagErr &  Error on NUV magnitude. \\

FUV\_Mag & FUV magnitude (AB). \\

FUV\_MagErr &  Error on FUV magnitude. \\

FIRST\_Name & Name of FIRST counterpart. \\

FIRST\_RA & RA of FIRST counterpart (J2000). \\

FIRST\_Dec & Dec of FIRST counterpart (J2000). \\

FIRST\_Flux & Integrated flux density at 1.4 GHz (mJy).\\

FIRST\_Err &  Error on the flux density, calculated by multiplying the integrated flux density by the ratio of the RMS to the peak flux (mJy).\\

HERS\_RA &  RA of {\it Herschel} counterpart from HeRS survey \citep[J2000;][]{viero}. \\

HERS\_Dec & Dec of {\it Herschel} counterpart from HeRS survey \citep[J2000;][]{viero}. \\

F250 &  Flux density at 250 $\mu$m (mJy).  \\

F250\_Err &  1$\sigma$ flux density error at 250 $\mu$m (mJy). \\

F350 & Flux density at 350 $\mu$m (mJy). \\

F350\_Err & 1$\sigma$ flux density error at 350 $\mu$m (mJy). \\

F500 & Flux density at 500 $\mu$m (mJy). \\

F500\_err & 1$\sigma$ flux error at 500 $\mu$m (mJy). \\

cp\_coord\_flag & Set to 0 if multi-wavelength counterpart coordinates are consistent within 2$^{\prime\prime}$ (SDSS, UKIDSS, VHS, FIRST) or 3$^{\prime\prime}$ ({\it WISE}, {\it GALEX}, {\it Herschel}); otherwise flag is set to 1. \\

\end{longtable}

\clearpage

\begin{table}
\caption{Additional Columns in {\it Chandra} Catalog}
\begin{tabular}{p{0.3\linewidth}p{0.6\linewidth}}
\hline
\noalign{\smallskip}\hline\noalign{\smallskip}
Column & Description \\
\noalign{\smallskip}\hline\noalign{\smallskip}

MSID &  {\it Chandra} Source Catalog unique identification number. \\

Soft\_Flux\_Error\_High & Higher bound on 0.5-2 keV flux (erg s$^{-1}$ cm$^{-2}$). If soft flux is zero, this represents upper limit. \\

Soft\_Flux\_Error\_Low &  Lower bound on 0.5-2 keV flux (erg s$^{-1}$ cm$^{-2}$). \\

Hard\_Flux\_Error\_High & Higher bound on 2-7 keV flux (erg s$^{-1}$ cm$^{-2}$). If hard flux is zero, this represents upper limit. \\

Hard\_Flux\_Error\_Low &  Lower bound on 2-7 keV flux (erg s$^{-1}$ cm$^{-2}$). \\

Full\_Flux\_Error\_High & Higher bound on 0.5-7 keV flux (erg s$^{-1}$ cm$^{-2}$). If full flux is zero, this represents upper limit. \\

Full\_Flux\_Error\_Low & Lower bound on 0.5-7 keV flux (erg s$^{-1}$ cm$^{-2}$). \\

In\_XMM &  Flag set to ``yes'' if source is detected in the archival and AO10 {\it XMM-Newton} catalog. The rec\_no of this associated {\it XMM-Newton} source is also given. \\

In\_XMM\_AO13 &  Flag set to ``yes'' if source is detected in the AO13 {\it XMM-Newton} catalog. The rec\_no of this associated {\it XMM-Newton} source is also given. \\
XMM\_archive\_cp\_flag &  Flag to indicate that a multi-wavelength counterpart is promoted into this catalog: the multi-wavelength association did not meet the MLE reliability threshold when matching to the {\it Chandra} catalog, but did meet this requirement for the same X-ray source in the {\it XMM-Newton} archival catalog. The number indicates from which catalog the promoted match is found: 1 - SDSS counterpart found but photometry rejected for failing quality control checks; 2 - SDSS;  3 - redshift; 4 - {\it WISE} counterpart found but rejected for failing quality control checks; 5 - {\it WISE}; 6 - UKIDSS; 7- VHS; 8 - {\it GALEX}. \\ 

XMM\_ao13\_cp\_flag &   Similar to the ``XMM\_archive\_cp\_flag'', but for matches from the archival {\it XMM-Newton} AO13 catalog. See ``XMM\_archive\_cp\_flag'' for more information. \\

\noalign{\smallskip}\hline\noalign{\smallskip}
\end{tabular}
\end{table}

\begin{table}
\caption{Additional Columns in {\it XMM-Newton} Catalog}
\begin{tabular}{p{0.3\linewidth}p{0.6\linewidth}}
\hline
\noalign{\smallskip}\hline\noalign{\smallskip}
Column & Description \\
\noalign{\smallskip}\hline\noalign{\smallskip}

Rec\_No & Unique identifying number for X-ray source. \\

Ext\_Flag &  Flag to indicate whether source was extended in one or more bands while being point-like in another band:  1 - extended in the soft band, 2 - extended in the full band, 3 - extended in the hard band, 4 - extended in the soft and full bands, 5 -  extended in the soft and hard bands, 6 - extended in the hard and full bands. If 0, then the source is point-like in all bands.\\

In\_XMM ({\it in AO13 catalog})  & Flag set to ``yes'' if source is detected in the archival and AO10 {\it XMM-Newton} catalog. The rec\_no of this associated {\it XMM-Newton} source is also given. \\

In\_XMM\_AO13 ({\it in archival and AO10 catalog}) & Flag set to ``yes'' if source is detected in the {\it XMM-Newton} AO13 catalog. The rec\_no of this associated {\it XMM-Newton} source is also given. \\

In\_Chandra & Flag set to ``yes'' if source is detected in the {\it Chandra} catalog. The MSID of this associated {\it Chandra} source is also given. \\

Soft\_Flux\_Err & Error on the 0.5-2 keV flux (erg s$^{-1}$ cm$^{-2}$).\\

Soft\_detml & Significance of the detection in the 0.5-2 keV band, where $det\_ml$=-ln$P_{\rm random}$. Users are cautioned to determine the flux significance necessary for their science goals before utilizing the reported flux. For reference, we only include objects where $det\_ml \geq 15$ in the Log$N$-Log$S$ relationship, and only report the luminosities for objects above this threshold. \\

Hard\_Flux\_Err & Error on the 2-10 keV flux (erg s$^{-1}$ cm$^{-2}$).\\

Hard\_detml & Significance of the detection in the 2-10 keV band. See ``soft\_detml'' for more information. \\

Full\_Flux\_Err & Error on the 0.5-10 keV flux (erg s$^{-1}$ cm$^{-2}$).\\

Full\_detml & Significance of the detection in the 0.5-10 keV band. See ``soft\_detml'' for more information. \\

XMM\_ao13\_cp\_flag ({\it in archival and AO10 catalog}) &  Flag to indicate that a multi-wavelength counterpart is promoted into this catalog: the multi-wavelength association did not meet the MLE reliability threshold when matching to the AO13 catalog, but did meet this requirement for the same X-ray source in the {\it XMM-Newton} archival catalog. The number indicates from which catalog the promoted match is found: 1 - SDSS counterpart found but photometry rejected for failing quality control checks; 2 - SDSS;  3 - redshift; 4 - {\it WISE} counterpart found but rejected for failing quality control checks; 5 - {\it WISE}; 6 - UKIDSS; 7 - VHS; 8 - {\it GALEX}. \\ 

XMM\_archive\_cp\_flag ({\it in AO13 catalog})  & Same as the ``XMM\_ao13\_cp\_flag'', but for promoted matches into the AO13 catalog from the archival and AO10 catalog. \\

Ch\_cp\_flag & Similar to the ``XMM\_ao13\_cp\_flag'', but for matches from the archival {\it Chandra} catalog. See ``XMM\_archive\_cp\_flag'' for more information. \\
\noalign{\smallskip}\hline\noalign{\smallskip}
\end{tabular}
\end{table}

\end{document}